\documentclass[aps,prd,preprint,eqsecnum,nofootinbib,groupedaddress,showpacs,floatfix,superscriptaddress]{revtex4}
\usepackage[dvips]{graphicx}
\usepackage{amssymb}
\newif\ifdraft
\drafttrue
\newif\ifpreprint
\preprinttrue

\def\nn{\nonumber}

\def\fig#1{fig.~{\ref{#1}}}

\def\sect#1{section~{\ref{#1}}}
\def\Sect#1{Section~{\ref{#1}}}

\def\eqn#1{eq.~(\ref{#1})}

\def\eqns#1#2{eqs.~(\ref{#1}) and~(\ref{#2})}

\def\tab#1{table~{\ref{#1}}}

\def\eps{\epsilon}

\def\Ord{{\cal O}}
\def\Res{\mathop{\rm Res}}

\def\spa#1.#2{\left\langle#1\,#2\right\rangle}
\def\spb#1.#2{\left[#1\,#2\right]}
\def\spash#1.#2{\spa{\smash{#1}}.{\smash{#2}}}
\def\spbsh#1.#2{\spb{\smash{#1}}.{\smash{#2}}}
\def\sand#1.#2.#3{%
\left\langle\smash{#1^{-}}{\vphantom1}\right|{#2}%
\left|\smash{#3^{-}}{\vphantom1}\right\rangle}
\def\sandp#1.#2.#3{%
\left\langle\smash{#1^{-}}{\vphantom1}\right|{#2}%
\left|\smash{#3^{+}}{\vphantom1}\right\rangle}
\def\sandpp#1.#2.#3{%
\left\langle\smash{#1^{+}}{\vphantom1}\right|{#2}%
\left|\smash{#3^{+}}{\vphantom1}\right\rangle}
\def\sandpm#1.#2.#3{%
\left\langle\smash{#1^{+}}{\vphantom1}\right|{#2}%
\left|\smash{#3^{-}}{\vphantom1}\right\rangle}
\def\sandmp#1.#2.#3{%
   \left\langle\smash{#1^{-}}{\vphantom1}\right|{#2}%
    \left|\smash{#3^{+}}{\vphantom1}\right\rangle}

\def\ketm#1{|\smash{#1}\vphantom{1}^-\rangle}
\def\ketp#1{|\smash{#1}\vphantom{1}^+\rangle}

\def\ssand#1.#2.#3{%
\left\langle\smash{#1}{\vphantom1}\right|{#2}%
\left|\smash{#3}{\vphantom1}\right]}
\def\ssandp#1.#2.#3{%
\left\langle\smash{#1}{\vphantom1}\right|{#2}%
\left|\smash{#3}{\vphantom1}\right\rangle}
\def\ssandpp#1.#2.#3{%
\left\langle\smash{#1}{\vphantom1}\right|{#2}%
\left|\smash{#3}{\vphantom1}\right\rangle}
\def\sketm#1{|\smash{#1}\vphantom{1}\rangle}
\def\sketp#1{|\smash{#1}\vphantom{1}]}

\def\proj{\flat}
\def\projdot#1.#2{k_{#1}^\proj\cdot k_{#2}^\proj}
\def\sandff#1.#2.#3{%
\left\langle\smash{#1^{\proj,-}}{\vphantom1}\right|{#2}%
\left|\smash{#3^{\proj,-}}{\vphantom1}\right\rangle}
\def\sandnf#1.#2.#3{%
\left\langle\smash{#1^{-}}{\vphantom1}\right|{#2}%
\left|\smash{#3^{\proj,-}}{\vphantom1}\right\rangle}
\def\sandfn#1.#2.#3{%
\left\langle\smash{#1^{\proj,-}}{\vphantom1}\right|{#2}%
\left|\smash{#3^{-}}{\vphantom1}\right\rangle}

\def\ketfm#1{|\smash{#1}\vphantom{1}^{\proj,-}\rangle}
\def\ketfp#1{|\smash{#1}\vphantom{1}^{\proj,+}\rangle}

\def\be{\begin{equation}}
\def\ee{\end{equation}}
\def\bea{\begin{eqnarray}}
\def\eea{\end{eqnarray}}

\newbox\charbox
\newbox\slabox
\def\s#1{{      % Feynman slash
        \setbox\charbox=\hbox{$#1$}
        \setbox\slabox=\hbox{$/$}
        \dimen\charbox=\ht\slabox
        \advance\dimen\charbox by -\dp\slabox
        \advance\dimen\charbox by -\ht\charbox
        \advance\dimen\charbox by \dp\charbox
        \divide\dimen\charbox by 2
        \raise-\dimen\charbox\hbox to \wd\charbox{\hss/\hss}
        \llap{$#1$}
}}

\def\etasl{\s{\eta}}

\def\onehalf{\frac12}
\def\Sol{{\cal S}}
\def\Global{{\cal G}}

\begin{document}

\hbox{
Saclay IPhT--T12/161$\hskip 0.55cm \null$
UUITP-22/12
}

\title{Two-Loop Maximal Unitarity with External Masses}

\author{Henrik Johansson}
\affiliation{Institut de Physique Th\'eorique, CEA--Saclay,
          F--91191 Gif-sur-Yvette cedex, France\\
          {\tt Henrik.Johansson@cea.fr}\\
          {\tt David.Kosower@cea.fr}}
\author{David~A.~Kosower}
\affiliation{Institut de Physique Th\'eorique, CEA--Saclay,
          F--91191 Gif-sur-Yvette cedex, France\\
          {\tt Henrik.Johansson@cea.fr}\\
          {\tt David.Kosower@cea.fr}}
\author{Kasper J.~Larsen}
\affiliation{Institut de Physique Th\'eorique, CEA--Saclay,
          F--91191 Gif-sur-Yvette cedex, France\\
          {\tt Henrik.Johansson@cea.fr}\\
          {\tt David.Kosower@cea.fr}}
\affiliation{School of Natural Sciences, \\ Institute for Advanced Study,
             Princeton, NJ 08540, USA\\
          {\tt Kasper.Larsen@cea.fr}}
\affiliation{Department of Physics and Astronomy, Uppsala
          University, SE--75108 Uppsala, Sweden\\
          {\tt Kasper.Larsen@cea.fr}}

\begin{abstract}
We extend the maximal unitarity method at two loops to double-box
basis integrals with up to three external massive legs.
We use consistency equations based on the requirement
that integrals of total derivatives vanish.  We obtain
unique formul\ae{} for the coefficients of the master double-box
integrals.
These formul\ae{} can
be used either analytically or numerically.
\end{abstract}


\maketitle

\section{Introduction}

A quantitative understanding of backgrounds at the LHC
to new-physics signals is important to direct searches
for physics beyond the Standard Model.
A similar understanding of signals of heavy Standard-Model particles
is important to detailed studies of their properties as another window
into new physics.  Such quantitative understanding
requires at least next-to-leading order (NLO) calculations in quantum
chromodynamics (QCD).  These calculations make use of a number
of ingredients beyond the tree-level amplitudes required for a leading-order
(LO) calculation:
real-emission corrections, with an additional emitted gluon, or a
gluon splitting into a quark--antiquark pair; and virtual one-loop
corrections, with a virtual gluon or virtual quark in a closed loop.
The required one-loop corrections are challenging with traditional
Feynman-diagram methods, and become considerably more difficult as the
number of final-state partons (gluons or quarks) grows.

Some subprocesses are absent at tree level, such as
the gluon fusion to
diphoton subprocess, $gg\rightarrow\gamma\gamma$, which is an
important background to measurements of the recently-discovered
new heavy boson (NHB)
at the LHC~\cite{AtlasHiggs,CMSHiggs}.  While these subprocesses
are nominally of higher order in the strong coupling $\alpha_s$,
the large gluon parton density at smaller
$x$ can compensate for this additional power, giving rise to
contributions to cross sections which are comparable to those from
tree-level quark-initiated
subprocesses~\cite{Ellis:1987xu,Berger:1983yi,Aurenche:1985yk}.
The production of electroweak boson pairs, $gg\rightarrow
Z\gamma,ZZ,W^+W^-$ also falls into this class.
Because the basic processes only arise from one-loop amplitudes,
computation of the NLO corrections requires two-loop
amplitudes~\cite{Bern:2002jx}
(as well as one-loop amplitudes with an additional parton, and
singular factors derived from one-loop amplitudes).

Two-loop amplitudes are also required for any studies beyond NLO.
Next-to-next-to-leading order (NNLO) fixed-order calculations stand at the
next frontier of precision QCD calculations.
The only existing fully-exclusive NNLO jet
calculations to date are for three-jet production in
electron--positron annihilation~\cite{NNLOThreeJet}.  These are
necessary to determine $\alpha_s$ to 1\% accuracy from jet data at
LEP~\cite{ThreeJetAlphaS}, competitively with other determinations.
At the LHC, NNLO calculations will be useful for determining an honest
theoretical uncertainty estimate on NLO calculations, for assessing scale
stability in multi-scale processes such as $W$+multi-jet production,
and will also be required for precision measurements of
the NHB or of new physics should it be discovered.

The unitarity
method~\cite{UnitarityMethod,Bern:1995db,Zqqgg,DdimensionalI,BCFUnitarity,OtherUnitarity,Bootstrap,BCFCutConstructible,BMST,OPP,OnShellReview,Forde,Badger,DdimensionalII,BFMassive,BergerFordeReview,Bern:2010qa}
has made many previously-inaccessible
one-loop calculations feasible.  Of particular note
are processes
with many partons in the final state.  The most recent
development, applying generalized unitarity, allows the
method to be applied either analytically or purely
numerically~\cite{EGK,BlackHatI,CutTools,MOPP,Rocket,BlackHatII,CutToolsHelac,Samurai,WPlus4,NGluon,MadLoop}.
The numerical formalisms underly recent software libraries and
programs that have been applied to LHC phenomenology.
In this approach,
the one-loop amplitude in QCD is written as a sum over a
set of basis integrals, with coefficients that are rational in
external spinors,
\begin{equation}
{\rm Amplitude} = \sum_{j\in {\rm Basis}}
  {\rm coefficient}_j \times {\rm Integral}_j +
{\rm Rational}\,.
\label{BasicEquation}
\end{equation}
The integral basis for amplitudes with massless internal lines contains
box, triangle, and bubble integrals in addition to purely rational
terms (dropping all terms of $\Ord(\eps)$ in the dimensional regulator).
  The coefficients
are calculated from products of tree amplitudes, typically by
performing contour integrals via discrete Fourier projection.

The unitarity method has also been applied to higher-loop
amplitudes.  Some prior applications made use
of `minimal' generalized unitarity, cutting just enough propagators to
break apart a higher-loop amplitude into a product of tree amplitudes.
Each cut is again a product of tree amplitudes, but because not
all possible propagators are cut, each generalized cut will
correspond to several integrals, and
algebra will be required to isolate specific integrals and
their coefficients.  This approach does not require a
predetermined general basis of integrals.
A number of calculations have been done this way, primarily
in the ${\cal N}=4$ supersymmetric gauge theory~\cite{Bern:1997nh,ABDK,
Bern:2005iz,Bern:2006vw,Bern:2006ew,Bern:2008ap,ArkaniHamed:2010kv,Kosower:2010yk,ArkaniHamed:2010gh,Bern:2011rj}, but including
several four-point calculations in QCD
and supersymmetric
theories with less-than-maximal supersymmetry~\cite{Bern:2000dn,Bern:2002tk,BernDeFreitasDixonTwoPhoton,Bern:2002zk,Bern:2003ck,TwoLoopSplitting,DeFreitas:2004tk}.
Furthermore, a number of recent multi-loop calculations in maximally supersymmetric gauge and gravity theories have used maximal cuts~\cite{Bern:2007ct, Bern:2008pv,LeadingSingularity,ArkaniHamed:2009dn,Bern:2010tq,Carrasco:2011mn,Carrasco:2011hw,Bern:2012uc}, without complete
localization of integrands.

In maximal generalized unitarity as we pursue it, one cuts as many propagators as
possible, and seeks to fully localize integrands onto global poles.
In principle, this allows one to isolate individual integrals
on the right-hand side of the higher-loop analog of \eqn{BasicEquation}.
In a previous paper~\cite{MaximalTwoLoopUnitarity},
two of the present authors showed how to extract
the coefficients of the two master double boxes using a
multi-dimensional contour
around global poles.   Each global pole corresponds to cutting all
propagators, and in addition seeking the poles of remaining degrees
of freedom (typically generated by the Jacobian from cutting the
propagators).  This approach may be viewed as a generalization to
two loops of the work of Britto, Cachazo, and Feng~\cite{BCFUnitarity},
and of Forde~\cite{Forde}.
In parallel, Mastrolia and Ossola~\cite{MastroliaOssola} and
Badger, Frellesvig, and~Zhang~\cite{BadgerFZ} have showed how to generalize the
approach of Ossola, Papadopoulos, and Pittau~\cite{OPP} to higher loops.
We refer to refs.~\cite{Mastrolia:2012an,Kleiss:2012yv,Feng:2012bm,Mastrolia:2012wf,Mastrolia:2012du}
for further recent developments within this approach. A recent paper by
Zhang~\cite{Zhang} adds tools required in such an approach
for reducing integrands to a basis of monomials.

In this paper, we continue the maximal-unitarity approach of
ref.~\cite{MaximalTwoLoopUnitarity}, taking advantage of recent
work~\cite{Caron-HuotLarsen}
by Caron-Huot and one of the present authors showing that the set of
distinct global poles is smaller than previously assumed.
Higher-loop
amplitudes can be written in a similar form to those at
one loop~(\ref{BasicEquation}),
as a sum over an integral basis~\cite{TwoLoopBasis},
along with possible rational terms.
(At higher loops, the coefficients of the basis
integrals are no longer rational functions
of the external spinors alone, but will depend explicitly
on the dimensional regulator $\eps$.)
Furthermore, at two and higher loops, the number of master
integrals for a given topology may depend on the number and
arrangement of external masses~\cite{TwoLoopBasis}.

As in ref.~\cite{MaximalTwoLoopUnitarity},
we use the equations relating
generic tensor integrals to basis (or master) integrals in order to ensure the
consistency and completeness of the choice of contours.
We obtain a set of unique projectors (or master contours)
to compute the coefficients of double boxes with one, two or three
external masses, one projector for each different master integral.
We again work only to leading order in $\eps$ in the coefficients.

The extraction
of the double-box coefficient bears a superficial similarity
to the procedure that would
be followed in the leading-singularity
approach~\cite{Octacut,LeadingSingularity},
but unlike the latter, manifestly ensures the consistency of
the extraction with respect to terms that vanish after integration.
Such terms inevitably arise when using the integration-by-parts (IBP)
approach~\cite{IBP,Laporta,GehrmannRemiddi,LIdependent,AIR,FIRE,Reduze,SmirnovPetukhov}
 in relating formally-irreducible tensor integrals to basis
integrals.
The extraction of higher-order terms in $\eps$ or the coefficients
of integrals with fewer propagators, both of which we leave to
future work, would also be different.

In this paper, we will focus on extracting the coefficients of
two-loop master integrals with massless internal lines. One can
consider the generalization to the case with massive internal
lines.  So long as there are some massless internal lines, so that
there is at least one chiral vertex, the integrand will still
have global poles, and we expect the approach described here to
generalize smoothly.  Masses on internal lines are also related
to the generalization to $D$-dimensional maximal unitarity,
which we again leave to future work.

In \sect{ParametrizingSection}, we present our parametrization of the
loop momenta.  In \sect{HeptacutSection}, we discuss the solutions
to the heptacut equations, and in \sect{GlobalPolesSection} we list
the global poles of the double boxes.  We discuss the constraint
equations in \sect{ConstraintSection}, and obtain our main
results for the projectors extracting integral coefficients
in \sect{ProjectorsSection}.  \Sect{ExampleSection} gives
examples of applying the technique described here, and provides
cross-checks with known results.  We give our conclusions in
\sect{ConclusionSection}.

\section{Parametrizing the Integrand}
\label{ParametrizingSection}

It will be convenient for our purposes to adopt a different parametrization
for the loop momenta from the one used in ref.~\cite{MaximalTwoLoopUnitarity}.
The parametrization used here has the virtue of keeping all expressions
manifestly rational throughout the derivation.
In this parametrization, we
will need spinors corresponding to massive legs.  To define appropriate
momenta, we make use of `mutually projected' momenta as in
the work of Ossola, Papadopoulos and Pittau~\cite{OPP}
and Forde~\cite{Forde} on the extraction of triangle and bubble
coefficients at one loop.  We use $(k_1,k_2)$ and $(k_3,k_4)$ as
mutually-projecting pairs, defining within each pair,
\begin{eqnarray}
k_{j,1}^{\proj,\mu} &=& k_{j,1}^\mu -
\frac{k_{j,1}^2}{2 k_{j,1}\cdot k_{j,2}^\proj} k_{j,2}^{\proj,\mu}
  \,,\nn\\
k_{j,2}^{\proj,\mu} &=& k_{j,2}^\mu -
 \frac{k_{j,2}^2}{2 k_{j,2}\cdot k_{j,1}^\proj} k_{j,1}^{\proj,\mu}
  \,.
\label{MutualProjection}
\end{eqnarray}
By construction, $k^\proj_{j,1}$ and $k^\proj_{j,2}$ are massless momenta.

\def\ibar{{\bar\imath}}
Define
\begin{equation}
\rho_{j,i} \equiv \frac{k_{j,i}^2}{2 k_{j,i}\cdot k_{j,\ibar}}\,,
\end{equation}
where $i$ denotes one momentum in the pair, and $\ibar$ the other.

We note that
\begin{equation}
k_{j,1}\cdot k_{j,2}^\proj
= k_{j,1}^\proj\cdot k_{j,2}
= k_{j,1}^\proj\cdot k_{j,2}^\proj\,.
\end{equation}
Define
\begin{equation}
\gamma_{j,12} \equiv 2 k_{j,1}^\proj\cdot k_{j,2}^\proj\,;
\end{equation}
we can solve for it to obtain,
\begin{equation}
\gamma_{j,12}^\pm =
  k_{j,1}\cdot k_{j,2} \pm
  \big[ (k_{j,1}\cdot k_{j,2})^2 - k_{j,1}^2 k_{j,2}^2\big]^{1/2}\,.
\end{equation}
If either momentum $k_{j,i}$ is massless, only one solution survives,
$\gamma_{j,12} = 2 k_{j,1}\cdot k_{j,2}$\,.

With two mutually-projecting pairs, we may have two independent
$\gamma$s, $\gamma_{12}$ for the $(k_1,k_2)$ pair, and
$\gamma_{34}$ for the $(k_3,k_4)$ pair.
In general, we expect to express our final results in terms of the
independent invariants $s_{12}$ and $s_{14}$, the nonzero masses
amongst $\{m_1,m_2,m_3,m_4\}$,
along with $\gamma_{12}$ (if $m_1\neq0\neq m_2$),
and $\gamma_{34}$ (if $m_3\neq0\neq m_4$).

Inverting \eqn{MutualProjection} we obtain the massless momenta
\begin{equation}
k_{j,i}^{\proj,\mu} = (1- \rho_{j,1} \rho_{j,2})^{-1}(k_{j,i}^{\mu} - \rho_{j,i} \,k_{j,\ibar}^{\mu})\,.
\end{equation}

\def\tlambdap{{\tilde\lambda'}{}}
We adopt the following parametrization for the
on-shell double-box loop momenta,
\begin{eqnarray}
\ell_1^\mu &=&
\onehalf\sand{\lambda_1}.{\mu}.{\tlambdap_1}
=\onehalf\ssand{\lambda_1}.{\mu}.{\tlambdap_1}
\,,\nn\\
\ell_2^\mu &=&
\onehalf\sand{\lambda_2}.{\mu}.{\tlambdap_2}
=\onehalf\ssand{\lambda_2}.{\mu}.{\tlambdap_2}\,.
\label{SpinorParametrization}
\end{eqnarray}
This form ensures that $\ell_1^2 = 0 = \ell_2^2$.
We write the various loop
spinors in terms of the spinors corresponding to $(k_{1}^\proj,k_{2}^\proj)$ for
$\ell_1$, and the spinors corresponding to $(k_{3}^\proj,k_{4}^\proj)$ for
$\ell_2$:
\begin{eqnarray}
\ketp{\lambda_1} &=& \xi_1 \ketfp{1}
   + \xi_2 \frac{\spash{4^\proj}.{1^\proj}}
                {\spash{4^\proj}.{2^\proj}} \ketfp{2}\,,\nn\\
\ketm{\tlambdap_1} &=& \xi'_1 \ketfm{1}
   + \xi'_2 \frac{\spbsh{4^\proj}.{1^\proj}}
                 {\spbsh{4^\proj}.{2^\proj}} \ketfm{2}\,,\nn\\
\ketp{\lambda_2} &=& \xi_3 \frac{\spash{1^\proj}.{4^\proj}}
            {\spash{1^\proj}.{3^\proj}} \ketfp{3} + \xi_4 \ketfp{4}\,,
\label{BasicSpinorDefinitions}\\
\ketm{\tlambdap_2} &=& \xi'_3 \frac{\spbsh{1^\proj}.{4^\proj}}
            {\spbsh{1^\proj}.{3^\proj}} \ketfm{3} + \xi'_4 \ketfm{4}\,.\nn
\end{eqnarray}
Without loss of generality, we can set $\xi_1 = 1 = \xi_4$.
We will use the simplified notation
$\sketm{j} \equiv \ketp{j}$ and
$\sketp{j} \equiv \ketm{j}$ in later sections of the paper.

\section{Heptacut Equations}
\label{HeptacutSection}
\vskip 10pt

\def\DBox{P^{**}_{2,2}}

\begin{figure}[!h]
\begin{center}
\includegraphics[angle=0, width=0.4\textwidth]{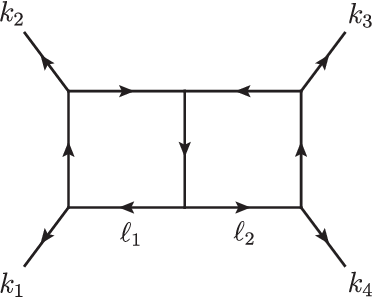}
\end{center}
\caption{The double-box integral $\DBox$.}
\label{PlanarTwoLoopDoubleBoxFigure}
\end{figure}

We focus on the planar double-box integral, shown
in~\fig{PlanarTwoLoopDoubleBoxFigure},
which is given by,
\begin{equation}
\DBox =
\int {d^{D} \ell_1\over (2\pi)^D} \hspace{0.5mm} {d^{D} \ell_2\over(2\pi)^D}
\hspace{0.7mm} \frac{1}{\ell_1^2(\ell_1 - k_1)^2
(\ell_1 - K_{12})^2(\ell_1 + \ell_2)^2\ell_2^2
(\ell_2 - k_4)^2(\ell_2 - K_{34})^2} \, ,
\label{eq:double_box_integral_def}
\end{equation}
where $K_{i\cdots j} \equiv k_i+\cdots+k_j$, and the notation
follows ref.~\cite{TwoLoopBasis}.  We allow up to three of the
external legs to be massive, leaving the four-mass case
for future investigation.  We leave the Feynman $i\varepsilon$
prescription implicit.

The heptacut equations correspond to cutting all seven propagators,
that is setting all denominator factors in eq.~(\ref{eq:double_box_integral_def})
to zero simultaneously.  Imposing the heptacut
can be implemented by {\it replacing\/} the original contour
of integration for the loop momenta, running along the real axis for
each component of $\ell_{1,2}^\mu$, by a seven-torus encircling the
solution surface(s) for the vanishing of these denominator factors.
Two of these equations are already solved by the parametrization in
\eqn{SpinorParametrization}.
Using this parametrization,
the remaining heptacut equations involving only $\ell_1$ take the form,
\begin{eqnarray}
(\ell_1-k_1)^2 &=& 0 ~~\Rightarrow\nn\\
m_1^2 &=& \biggl(\rho_{1,1} \xi'_1 + \frac{\projdot1.4}{\projdot2.4} \xi_2\xi'_2
                \biggr)\gamma_{12}\,,\nn\\
(\ell_1-K_{12})^2 &=& 0 ~~\Rightarrow
  \label{HeptacutOneRest} \\
s_{12}-m_1^2 &=& \biggl(\xi'_1+\rho_{1,2}\frac{\projdot1.4}{\projdot2.4}\xi_2\xi'_2
                \biggr)\gamma_{12}\,.\nn
\end{eqnarray}

\def\xib{\bar\xi}
In the general case ($m_1 \neq 0 \neq m_2$), these
equations have the solution,
\begin{eqnarray}
\xi'_1 &=& \frac{\gamma_{12} s_{12}-(\gamma_{12}+m_2^2) m_1^2}
                {\gamma_{12}^2-m_1^2 m_2^2} \equiv \xib'_1\,,\nn\\
\xi'_2 &=& -\frac{m_1^2 \bigl( s_{12}-\gamma_{12}-m_1^2\bigr)
               \projdot2.4}
                 {\xi_2 \bigl(\gamma_{12}^2-m_1^2 m_2^2\bigr)
                  \projdot1.4} \equiv \frac{\xib'_2}{\xi_2}\,.
\label{GeneralISolutions}
\end{eqnarray}
There are similar solutions for the equations involving only $\ell_2$.

The first of these solutions has the correct limits when either
$m_1 \rightarrow 0$ or $m_2 \rightarrow 0$ (or both); the second solution,
in contrast, is replaced by the pair of solutions,
\begin{equation}
\xi_2 = 0\,,\quad \xi'_2 {\rm\ free\/};\qquad
 {\rm\ or\ } \qquad \xi'_2 = 0\,,\quad \xi_2 {\rm\ free\/}\,.
\end{equation}
More explicitly, when $m_1 = 0$ but $m_2 \neq 0$,
\begin{equation}
\xi'_1 = {s_{12}\over\gamma_{12}} = 1+\rho_{1,2}\,;
\end{equation}
when $m_2 = 0$ (whether $m_1$ vanishes or not),
\begin{equation}
\xi'_1 = 1\,.
\end{equation}

The last heptacut equation is $\ell_1\cdot \ell_2=0$, which takes the form,
\begin{eqnarray}
0 &=& -\frac{1}{2} \spash{1^\proj}.{4^\proj}\spbsh{1^\proj}.{4^\proj}
\left(1  + \xi_2 +\xi_3
    +\xi_2\xi_3 {\spash{1^\proj}.{4^\proj}\spash{2^\proj}.{3^\proj}
                 \over\spash{2^\proj}.{4^\proj}\spash{1^\proj}.{3^\proj}}
\right)\nn\\
&&\hskip 7mm\times\left(
\xib'_1 \xib'_4  + {\xib'_2 \xib'_4\over\xi_2} +{\xib'_1\xib'_3\over\xi_3}
    +{\xib'_2\xib'_3\over \xi_2\xi_3}
        {\spbsh{1^\proj}.{4^\proj}\spbsh{2^\proj}.{3^\proj}
                 \over\spbsh{2^\proj}.{4^\proj}\spbsh{1^\proj}.{3^\proj}}
\right)\,.
\label{LastEquation}
\end{eqnarray}
\def\taub{{\bar\tau}}
This equation has two solutions corresponding to the two factors,
\begin{equation}
\xi_3 = \left\{\begin{array}{c}
 \displaystyle
  -\frac{1+\xi_2}{1+\tau \xi_2}\,,\\[3mm]
 \displaystyle
  -\frac{(\xi_2 \xib'_1 + \taub \xib'_2)\xib'_3}
        {(\xi_2 \xib'_1+\xib'_2) \xib'_4}\,,\end{array}\right.
\label{LastEquationGeneralSolution}
\end{equation}
where
\begin{eqnarray}
\tau &\equiv& \frac{\spash{1^\proj}.{4^\proj}\spash{2^\proj}.{3^\proj}}
                   {\spash{2^\proj}.{4^\proj}\spash{1^\proj}.{3^\proj}}\,,
\nn\\
\taub &\equiv& \frac{\spbsh{1^\proj}.{4^\proj}\spbsh{2^\proj}.{3^\proj}}
                    {\spbsh{2^\proj}.{4^\proj}\spbsh{1^\proj}.{3^\proj}}\,.
\end{eqnarray}
In fact,
\begin{eqnarray}
\tau = \taub &=&
-\frac{\gamma_{34} \bigl(\gamma_{12}+m_1^2\bigr)}
       {\bigl(\gamma_{34}+m_3^2\bigr)
            \bigl[(\gamma_{12}\gamma_{34}-m_1^2 m_3^2)
                  (\gamma_{12}\gamma_{34}-m_2^2 m_4^2)
                  +\gamma_{12}\gamma_{34} s_{12} s_{14}\bigr]}\nn\\
&&\hskip 5mm\times
      \Big[ (\gamma_{12}+m_2^2)(\gamma_{34}+m_3^2) (m_2^2+m_3^2-s_{14})
            + 2 m_2^2 m_3^2 s_{12}
             \\
&&\hskip 5mm\hphantom{\bigl[]}
            +(\gamma_{12}+m_2^2) m_3^2 (m_1^2-m_2^2-s_{12})
            +(\gamma_{34}+m_3^2) m_2^2 (m_4^2-m_3^2-s_{12})
      \Big]\,.\nn
\end{eqnarray}

As in ref.~\cite{Caron-HuotLarsen}, we must consider three different
classes of solutions, corresponding to different configurations of
external masses.

%%%%%%%%% FIGURE %%%%%%%%%%%%%%%%%%
\begin{figure}[th]
\begin{center}
\includegraphics[width=0.35\textwidth]{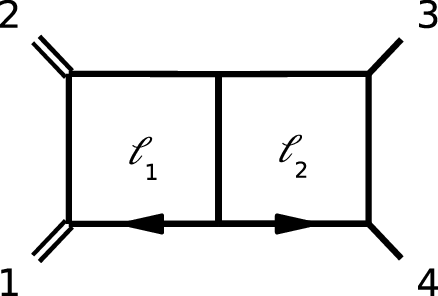}
\end{center}
\caption{\small The short-side two-mass double box.  Single lines indicate
massless legs, and doubled lines massive legs (here the two legs on the
left-hand side).}
\label{ShortSideTwoMassDoubleBoxFigure}
\end{figure}
%%%%%%%%%%%%%%%%%%%%%%%%%%%%%%%%

%%%%%%%%% FIGURE %%%%%%%%%%%%%%%%%%
\begin{figure}[th]
\begin{center}
\includegraphics[width=0.35\textwidth]{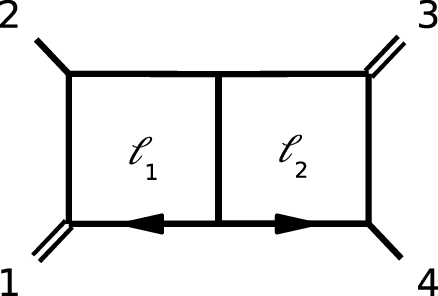}
\hskip 1.5cm
\includegraphics[width=0.35\textwidth]{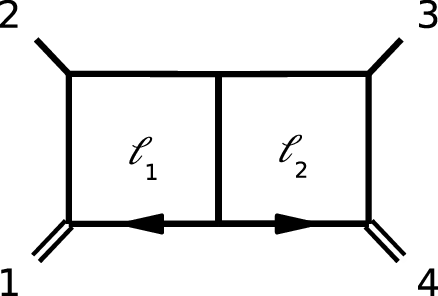}\\
\null\hskip 26mm\hbox to 0.3\textwidth{(i)}\hskip 1cm
\hbox to 3cm{(ii)}
\end{center}
\caption{\small Other two-mass double-boxes: (i) the diagonal one (ii)
  the long-side one.  Single lines indicate massless legs, and doubled
  lines massive legs (in (i), the lower-left and upper-right legs, and
  in (ii), the two lower legs).}
\label{OtherTwoMassDoubleBoxesFigure}
\end{figure}
%%%%%%%%%%%%%%%%%%%%%%%%%%%%%%%%

In considering massless limits, there are three classes to
consider: (a) when neither pair $(1,2)$ or $(3,4)$ contains a massless
momentum; (b) when exactly one pair contains a massless momentum; (c)
when both pairs contain massless momenta.  The first class corresponds
to the four-mass double box; the second class, to the three-mass double
box and the short-side two-mass one, shown in
\fig{ShortSideTwoMassDoubleBoxFigure};
the last class, to the remaining
configurations --- diagonal and long-side two-mass double boxes, shown
in \fig{OtherTwoMassDoubleBoxesFigure},
along
with the one-mass and massless ones.  Class (a) is given by the solutions
above, but we will not consider it any further in the present
paper.

In class (b), it suffices to take the massless momentum to
be either $k_3$ or $k_4$ (or both).  There are two solutions to the $\ell_2$
equations to consider, $\xi_3 = 0$ and $\xi'_3 = 0$.  For the first,
the last heptacut equation becomes,
\begin{equation}
0 = \spash{1^\proj}.{4^\proj}\spbsh{1^\proj}.{4^\proj}
\bigl(1  + \xi_2\bigr)
\left(
\xib'_1 \xib'_4  + {\xib'_2 \xib'_4\over\xi_2} +{\xib'_1\xi'_3}
    +\taub{\xib'_2\xi'_3\over \xi_2}
\right)
\label{LastEquationIIa}
\end{equation}
This equation again has two solutions, $\xi_2 = -1$ (with $\xi'_3$ being
identified as the remaining variable $z$), and the other having,
\begin{equation}
\xi'_3 = -\frac{\xib'_4(\xib'_1\xi_2 + \xib'_2)}
               {\xib'_1\xi_2 + \taub\xib'_2}
\end{equation}
with $\xi_2 = z$.
The second solution to the $\ell_2$ heptacut equations has
$\xi'_3 = 0$, whereupon the last heptacut equation becomes,
\begin{equation}
0 = \spash{1^\proj}.{4^\proj}\spbsh{1^\proj}.{4^\proj}
\bigl(1  + \xi_2 +\xi_3 +\tau \xi_2\xi_3
\bigr) \left(
\xib'_1\xib'_4  + {\xib'_2\xib'_4\over\xi_2}
\right)\,,
\label{LastEquationIIb}
\end{equation}
which also has two solutions,
\begin{equation}
\xi_2 = -{\xib'_2\over\xib'_1}
\end{equation}
(with $\xi_3=z$), and
\begin{equation}
\xi_3 = -{1+\xi_2\over 1+\tau\xi_2}
\end{equation}
(with $\xi_2 = z$).
Overall in class (b), we have four solutions to the heptacut equations,
in agreement with the discussion in ref.~\cite{Caron-HuotLarsen},
with $\xi'_1=\xib'_1$, $\xi'_4=\xib'_4$ in all, and the four
variables $(\xi_2,\xi'_2,\xi_3,\xi'_3)$ taking on the following values,
\begin{eqnarray}
\Sol_1:&&
\left(-\frac{\xib'_2}{\xib'_1},-\xib'_1, z,0
\right)\,,
\nn\\
\Sol_2:&&
\left(z,\frac{\xib'_2}z,-\frac{1+z}{1+\tau z} ,0
\right)\,,
\nn\\
\Sol_3:&&
\left( -1,-\xib'_2,0,z\right)\,,
\label{ClassBSolutions}\\
\Sol_4:&&
\left( z,\frac{\xib'_2}z,0,
-\frac{\xib'_4(\xib'_1 z + \xib'_2)}
               {\xib'_1 z + \taub\xib'_2}\right)
\,.\nn
\end{eqnarray}

In class (c), we have four solutions to the pure-$\ell_{1,2}$
equations to consider,
\begin{eqnarray}
\xi_2 = 0 &{\rm\ and\ \/}& \xi_3 = 0\,,\nn\\
\xi_2 = 0 &{\rm\ and\ \/}& \xi'_3 = 0\,,\nn\\
\xi'_2 = 0 &{\rm\ and\ \/}& \xi_3 = 0\,, \label{ClassC}\\
\xi'_2 = 0 &{\rm\ and\ \/}& \xi'_3 = 0\,.\nn
\end{eqnarray}
In the first of these solutions, the last heptacut equation simplifies
to,
\begin{equation}
0 = \spash{1^\proj}.{4^\proj}\spbsh{1^\proj}.{4^\proj}
\bigl(
\xib'_1\xib'_4  + \xi'_2\xib'_4 +\xib'_1\xi'_3
    +\taub \xi'_2\xi'_3
\bigr)\,,
\label{LastEquationIIIa}
\end{equation}
which has the solution,
\begin{equation}
\xi'_3 = -{\xib'_4 (\xib'_1+\xi'_2)\over \xib'_1+\taub\xi'_2}
\end{equation}
(with $\xi'_2 = z$).
In the second solution in \eqn{ClassC},
the last heptacut equation simplifies to,
\begin{equation}
0 = \spash{1^\proj}.{4^\proj}\spbsh{1^\proj}.{4^\proj}
\bigl(1  +\xi_3 \bigr)
\bigl( \xib'_1\xib'_4  + \xi'_2\xib'_4\bigr)
\label{LastEquationIIIb}
\end{equation}
which has two solutions, $\xi_3 = -1$ (with $\xi'_2 = z$),
and $\xi'_2 = -\xib'_1$ (with $\xi_3 = z$).

In the third solution in \eqn{ClassC},
the last heptacut equation simplifies to,
\begin{equation}
0 = \spash{1^\proj}.{4^\proj}\spbsh{1^\proj}.{4^\proj}
\bigl(1  + \xi_2
\bigr) \bigl(
\xib'_1\xib'_4  +\xib'_1\xi'_3
\bigr)\,,
\label{LastEquationIIIc}
\end{equation}
which also has two solutions, $\xi_2 = -1$ (with $\xi'_3=z$),
and $\xi'_3 = -\xib'_4$ (with $\xi_2 = z$).

In the last solution to \eqn{ClassC},
the last heptacut equation simplifies to,
\begin{equation}
0 = \spash{1^\proj}.{4^\proj}\spbsh{1^\proj}.{4^\proj} \xib'_1\xib'_4
\bigl(1  + \xi_2 +\xi_3
    +\tau\xi_2\xi_3
\bigr)\,,
\label{LastEquationIIId}
\end{equation}
which has a single solution,
\begin{equation}
\xi_3 = -{1+\xi_2\over 1+\tau\xi_2}
\end{equation}
(with $\xi_2 = z$).

Overall in class (c), we have six solutions to the heptacut equations,
in agreement with refs.~\cite{MaximalTwoLoopUnitarity,Caron-HuotLarsen},
with $\xi'_1=\xib'_1$, $\xi'_4=\xib'_4$ in all, and the four
variables $(\xi_2,\xi'_2,\xi_3,\xi'_3)$ taking on the following values,
\begin{eqnarray}
\Sol_1:&&
\left(0,-\xib'_1,z,0
\right)\,,
\nn\\
\Sol_2:&&
\left(0,z,-1,0
\right)\,,
\nn\\
\Sol_3:&&
\left(-1,0,0,z
\right)\,,
\nn\\
\Sol_4:&&
\left(z,0,0,-\xib'_4
\right)\,,
\label{ClassCSolutions}\\
\Sol_5:&&
\left(z,0,-\frac{1+z}{1+\tau z},0
\right)\,,
\nn\\
\Sol_6:&&
\left(0,z,0,-\frac{\xib'_4 (\xib'_1+z)}{\xib'_1+\taub z}
\right)\,.
\nn
\end{eqnarray}
% list solutions directly for \ell_i?

\section{Global Poles}
\label{GlobalPolesSection}

As we take the cuts strictly in four dimensions, the
two loop momenta have a total of eight degrees of freedom.
After changing variables to the various $\xi_i$ and $\xi'_i$,
replacing the original contours of integration with tori encircling
the locations of the zeros of the denominator of the integrand
localizes it on the solutions of the heptacut equations.
We found the solutions to these equations in the previous section.
In the all-massless, one-mass, diagonal two-mass, and long-side two-mass
double boxes, corresponding to class (c), there are six independent
solutions.  In the short-side two-mass and in the three-mass double boxes,
corresponding to class (b), there are four independent solutions.

In either case, the heptacut equations fix only seven of the eight
degrees of freedom.  Performing the corresponding contour integrals
gives rise to an inverse Jacobian from dia\-gonalizing
and linearizing the denominator in the neighborhood of the simultaneous
zeros in each of the denominator factors.  In all cases, this inverse
Jacobian also has poles in the remaining degree of freedom $z$.  In certain
solutions, powers of one or the other of the loop momenta will also
have poles in $z$.  These poles, along with the Jacobian poles, make it possible to choose contours for the $z$
integration which further localizes the integral.
As we shall see, this in turn makes it
possible to distinguish between the various master integrals.

There are two distinct Jacobians that arise.  One emerges from the
change of variables
$\ell_{1,2}^\mu$ to the $\xi$ parameters.  This Jacobian appears in
the numerator.
The other is the inverse Jacobian that arises
from performing the heptacut contour integrals.  Because the parametrization
in eqs.~(\ref{SpinorParametrization})-(\ref{BasicSpinorDefinitions}) automatically ensures that $\ell_1^2=0=\ell_2^2$,
it is not suitable for computing these Jacobians.  As a remedy, add
an additional vector to the original
parametrization~(\ref{SpinorParametrization}),
\begin{eqnarray}
\ell_1^\mu &=& \frac{1}{2} \sand{\lambda_1}.{\mu}.{\tlambdap_1}
+\zeta_1\eta_1^{\mu}\,,\nn\\
\ell_2^\mu &=& \frac{1}{2}\sand{\lambda_2}.{\mu}.{\tlambdap_2}
+\zeta_2\eta_2^{\mu}\,,
\label{FullSpinorParametrization}
\end{eqnarray}
where $\zeta_i$ are complex numbers, and the $\eta_i$ are null vectors
satisfying $\etasl_1\ketp{\lambda_1} \neq 0\neq \etasl_1\ketm{\tlambdap_1}$
and $\etasl_2\ketp{\lambda_2} \neq 0\neq \etasl_2\ketm{\tlambdap_2}$.

In both class~(b) and~(c),
we may choose $\eta_1 = k_2^\proj$ and $\eta_2=k_3^\proj$.
This is a suitable choice so long as $\xi'_1\neq 0$,
so that the non-collinearity condition above is satisfied.
For generic momenta, this holds.
With the additional term, the equations $\ell_1^2 = 0 = \ell_2^2$ become,
\begin{eqnarray}
0 &=& \gamma_{12} \xi'_1\zeta_1 \,,\nn\\
0 &=& \gamma_{34} \xi'_4 \zeta_2\,.
\end{eqnarray}
The requirement that $\xi'_1\neq 0\neq\xi'_4$ leaves
only the obvious solutions, $\zeta_1 = 0$ and $\zeta_2 = 0$,
respectively, to these equations.

The Jacobian for the change of variables is,
\begin{equation}
J_j = \det_{\mu,i} {\partial\ell^\mu_j\over\partial v_{j,i}}\,,
\end{equation}
where $v_{j,1} = \zeta_j$, $v_{1,2} = \xi'_1$, $v_{2,2} = \xi'_4$,
$v_{1,3} = \xi_2$, $v_{2,3} = \xi_3$, $v_{1,4} = \xi'_2$,
and $v_{2,4} = \xi'_3$.

In all solutions in both classes~(b) and~(c), these Jacobians take the form,
\begin{eqnarray}
J_1 &=&
i \frac{\gamma _{12}^2 \projdot1.4 \xi'_1}{4 (\projdot2.4)}
\,,\nn\\
J_2 &=&
i \frac{\gamma _{34}^2 \projdot1.4 \xi'_4}{4 (\projdot1.3)}
\,.
\label{ChangeOfVariableJacobian}
\end{eqnarray}

\def\hept{{\rm h}}
The Jacobian that arises
from performing the heptacut contour integrals (which will
appear in the denominator) is,
\begin{equation}
J_\hept = \det_{i,j} {\partial P_i\over\partial w_j}\,,
\end{equation}
where $P_i$ is the $i$-th propagator denominator, and the set $\{w_j\}$
are the seven variables (out of the eight $\xi'_1$, $\xi_2$, $\xi'_2$,
$\xi_3$, $\xi'_3$, $\xi'_4$, $\zeta_1$, $\zeta_2$) frozen by the
contour integrations.

This Jacobian is not the same for all solutions $\Sol_i$.
As discussed earlier, it will appear in the denominator;
the overall Jacobian appearing after performing the heptacut contour
integrals is,
\def\ji{J_{\oint_{}^{}}}
\begin{equation}
\ji^{(0)} = \frac{J_1 J_2}{J_\hept}\,.
\end{equation}
For our purposes, it will be convenient to factor out an overall
prefactor which would otherwise cancel in final formul\ae{} for
integral coefficients, defining the reduced Jacobian for
class~(b) as,
\begin{equation}
\ji = \ji^{(0)} \frac{(\xib'_1-\xib'_2)\xib'_4}{\gamma_*}\,,
\end{equation}
and for class~(c) as,
\begin{equation}
\ji = \ji^{(0)} \frac{\xib'_1\xib'_4}{\gamma_*}\,,
\end{equation}
where
\begin{equation}
\gamma_* \equiv {\gamma_{12}\gamma_{34}
                 \over 32 \projdot1.4\,
              (\gamma_{12}^2-m_1^2 m_2^2)(\gamma_{34}^2-m_3^2 m_4^2)}\,.
\end{equation}

%%%%%%%%% FIGURE %%%%%%%%%%%%%%%%%%
\begin{figure}[th]
\begin{center}
\includegraphics[width=0.5\textwidth]{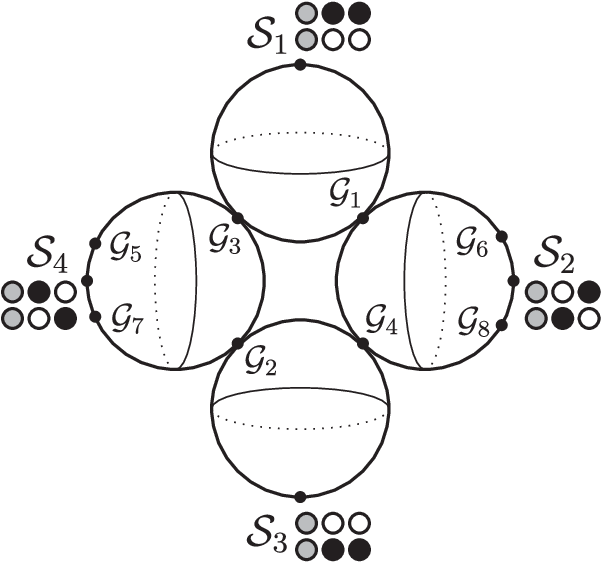}
\end{center}
\caption{\small A representation of the solution space for the class~(b)
heptacut equations, showing the four independent solutions $\Sol_i$,
and the locations of the eight global poles $\Global_j$.
 The small white, black and gray blobs indicate the pattern of chiral, antichiral and nonchiral
  kinematics, respectively, at the vertices of a double-box heptacut.}
\label{ClassBSolutionsFigure}
\end{figure}
%%%%%%%%%%%%%%%%%%%%%%%%%%%%%%%%

Within class~(b), we then find,
\begin{equation}
\ji =
\left\{\begin{array}{ll}
\displaystyle
\ji^{({\rm b},1)} \equiv
-\frac{\xib'_1-\xib'_2}
  {z \left(-\tau  z \xib'_2+z \xib'_1+\xib'_1-\xib'_2\right)}\,,
&{\rm for\ solution\/}~ \Sol_1\,,\\[4mm]
\displaystyle
\ji^{({\rm b},2)} \equiv
-\frac{\xib'_1-\xib'_2}{(z+1) \left(z \xib'_1+\xib'_2\right)}\,,
&{\rm for\ solution\/}~ \Sol_2\,,\\[4mm]
\displaystyle
\ji^{({\rm b},3)} \equiv
-\frac{(\xib'_1-\xib'_2)\xib'_4}{z \left(-\taub z \xib'_2+z \xib'_1+\xib'_4 \xib'_1-\xib'_2 \xib'_4\right)}\,,
&{\rm for\ solution\/}~ \Sol_3\,,\\[4mm]
\displaystyle
\ji^{({\rm b},4)} \equiv
\frac{\xib'_1-\xib'_2}{(z+1) \left(z \xib'_1+\xib'_2\right)}\,,
&{\rm for\ solution\/}~ \Sol_4\,.
\end{array}\right.
\end{equation}

The first of these Jacobians has poles at,
\begin{eqnarray}
z &=& 0\,,\nn\\
z &=& -\frac{\xib'_1-\xib'_2}{\xib'_1-\tau\xib'_2}\,;
\label{ClassBJacobianPoleIII}
\end{eqnarray}
the second and fourth at,
\begin{eqnarray}
z &=& -1\,,\nn\\
z &=& -\frac{\xib'_2}{\xib'_1}\,;
\label{ClassBJacobianPoleII}
\end{eqnarray}
and the third at,
\begin{eqnarray}
z &=& 0\,,\nn\\
z &=& -\frac{\xib'_4 (\xib'_1-\xib'_2)}{\xib'_1-\taub \xib'_2}\,.
\label{ClassBJacobianPoleI}
\end{eqnarray}

Integrals with powers of the loop momentum in the numerator will
have powers of the various coefficients $\xi_2,\xi'_2,\xi_3,\xi'_3$
once re-expressed in terms of our parametrization of loop
momenta~(\ref{SpinorParametrization}).  Some of these coefficients
also have poles in $z$, which leads to additional poles in the
integrand beyond those in the Jacobian above.  In each solution,
there is a pole at $z=\infty$. From
\eqn{ClassBSolutions}, we see that there are further poles at,
\begin{eqnarray}
z &=& 0\,,\nn\\
z &=& -\frac1{\tau}\,,
\label{ClassBSupplementaryPolesI}
\end{eqnarray}
for solution $\Sol_2$, and,
\begin{eqnarray}
z &=& 0\,,\nn\\
z &=& -\frac{\taub\xib'_2}{\xib'_1}\,,
\label{ClassBSupplementaryPolesII}
\end{eqnarray}
for solution $\Sol_4$.

Poles in $z$ complete the
poles in the other seven degrees of freedom to provide a \emph{global
pole}. Ultimately, we will be interested in residues at these global poles.
Not all poles within each solution are independent because the sum
of residues vanishes, and so we can remove one pole within each solution $\Sol_i$.
Furthermore, one might expect poles in $z$ in different solutions $\Sol_i$ to yield
distinct global poles; but this is not the case.
As pointed out~\cite{Caron-HuotLarsen} by Caron-Huot and one of the present authors,
global poles within one solution may in fact end up being identical
to global poles in another, in the sense of being located at the same values of
$\ell_{1,2}$. After removing such duplicates, one finds a total of
eight global poles.  At all eight, $\xi'_1=\xib'_1$,
$\xi'_4 = \xib'_4$, and $\zeta_1=0=\zeta_2$, while the values of
$(\xi_2,\xi'_2,\xi_3,\xi'_3)$ are as follows,
\begin{eqnarray}
\Global_1:
&&\left(-\frac{\xib'_2}{\xib'_1},-\xib'_1,\frac{\xib'_2-\xib'_1}{\xib'_1-\tau  \xib'_2},0\right)
\,,\nn\\
\Global_2:
&&\left(-1,-\xib'_2,0,-\frac{\left(\xib'_1-\xib'_2\right) \xib'_4}{\xib'_1-\taub \xib'_2}\right)
\,,\nn\\
\Global_3:
&&\left(-\frac{\xib'_2}{\xib'_1},-\xib'_1,0,0\right)
\,,\nn\\
\Global_4:
&&\left(-1,-\xib'_2,0,0\right)
\,,\label{ClassBGlobalPoles}\\
\Global_5:
&&\left(0,\infty,0,-\frac{\xib'_4}{\taub}\right)
\,,\nn\\
\Global_6:
&&\left(\infty,0,-\frac{1}{\tau},0\right)
\,,\nn\\
\Global_7:
&&\left(-\frac{\taub \xib'_2}{\xib'_1},-\frac{\xib'_1}{\taub},0,\infty\right)
\,,\nn\\
\Global_8:
&&\left(-\frac{1}{\tau },-\tau  \xib'_2,\infty,0\right)
\,.\nn
\end{eqnarray}
This corresponds to picking the two additional poles in eq.~(\ref{ClassBSupplementaryPolesII})
for solution $\Sol_4$, and the poles at $z=-1/\tau$ and $z=\infty$ for solution $\Sol_2$.

We obtain the same poles whether we pick $\gamma_{12} = \gamma_{12}^+$
or $\gamma_{12}^-$, so we can choose either.  In future numerical
applications, it may be more appropriate to average over the two choices.

In the poles where one parameter or another goes to infinity, we must
specify the order of limits carefully.

In global pole $\Global_5$, we should set,
\begin{equation}
\xi'_2= \frac{\xib'_2}{z}
\,,\, \xi _3= 0
\,,\, \xi'_3= -\frac{\xib'_4 \left(z \xib'_1+\xib'_2\right)}
                    {\taub \xib'_2+z \xib'_1}
\,,\, \xi _2= z
\,,
\end{equation}
followed by the limit $z\rightarrow 0$;
in global pole $\Global_6$, we should set,
\begin{equation}
\xi'_2= \frac{\xib'_2}{z}
\,,\, \xi'_3= 0
\,,\, \xi _3= -\frac{z+1}{\tau z+1}
\,,\, \xi _2= z
\,,
\end{equation}
followed by the limit $z\rightarrow \infty$;
in global pole $\Global_7$, we should set,
\begin{equation}
 \xi'_2= \frac{\xib'_2}{z}
\,,\, \xi _3= 0
\,,\, \xi'_3= -\frac{\xib'_4 \left(z \xib'_1+\xib'_2\right)}
                    {\taub \xib'_2+z \xib'_1}
\,,\, \xi _2= z
\,,\,
\,\end{equation}
followed by the limit $z\rightarrow -{\taub \xib'_2}/{\xib'_1}$;
and in global pole $\Global_8$, we should set,
\begin{equation}
\xi'_2= \frac{\xib'_2}{z}
\,,\, \xi'_3= 0
\,,\, \xi _3= -\frac{z+1}{\tau  z+1}
\,,\, \xi _2= z
\,,\,
\end{equation}
followed by the limit $z= -{1}/{\tau }$.
The solutions and global poles are depicted graphically in
\fig{ClassBSolutionsFigure}.

%%%%%%%%% FIGURE %%%%%%%%%%%%%%%%%%
\begin{figure}[th]
\begin{center}
\includegraphics[width=0.5\textwidth]{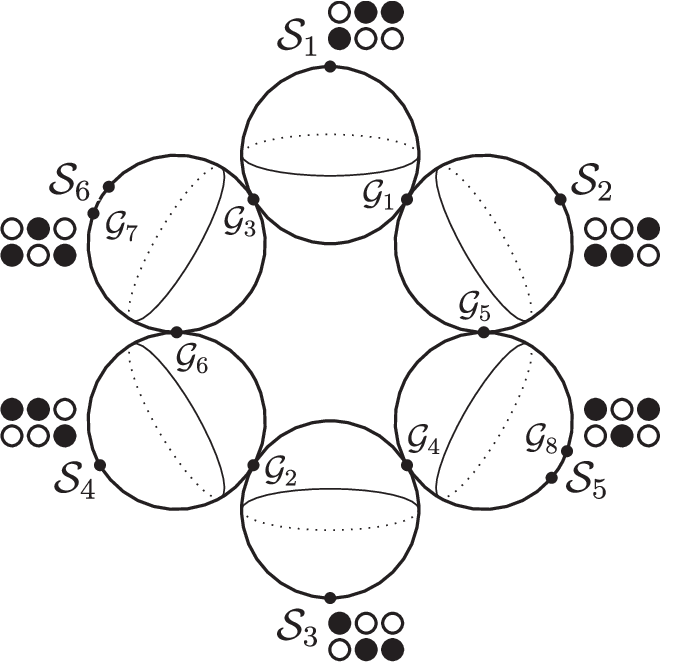}
\end{center}
\caption{\small A representation of the solution space for the class~(c)
heptacut equations, showing the six independent solutions $\Sol_i$,
and the locations of the eight global poles $\Global_j$.
The small white and black blobs indicate the pattern of chiral and antichiral kinematics,
respectively, at the vertices of a double-box heptacut.}
\label{ClassCSolutionsFigure}
\end{figure}
%%%%%%%%%%%%%%%%%%%%%%%%%%%%%%%%

Within class~(c) we find,
\begin{equation}
\ji =
\left\{\begin{array}{ll}
\displaystyle
\ji^{({\rm c},1)} \equiv
-\frac{1}{z (z+1)}\,,
&{\rm for\ solutions\/}~\Sol_{1,5}\,,\\[4mm]
\displaystyle
\ji^{({\rm c},2)} \equiv
\frac{\xib'_1}{z \left(\xib'_1+z\right)}\,,
&{\rm for\ solution\/}~\Sol_{2}\,,\\[4mm]
\displaystyle
\ji^{({\rm c},3)} \equiv
-\frac{\xib'_4}{z \left(\xib'_4+z\right)}\,,
&{\rm for\ solution\/}~\Sol_{3}\,,\\[4mm]
\displaystyle
\ji^{({\rm c},4)} \equiv
\frac{1}{z (z+1)}\,,
&{\rm for\ solution\/}~\Sol_4\,,\\[4mm]
\displaystyle
\ji^{({\rm c},5)} \equiv
-\frac{\xib'_1}{z \left(\xib'_1+z\right)}\,,
&{\rm for\ solution\/}~\Sol_{6}\,.
\end{array}\right.
\end{equation}

The first and fourth of these Jacobians have poles at,
\begin{eqnarray}
z &=& 0\,,\nn\\
z &=& -1\,;
\label{ClassCJacobianPoleIII}
\end{eqnarray}
the second and fifth at,
\begin{eqnarray}
z &=& 0\,,\nn\\
z &=& -\xib'_1\,;
\label{ClassCJacobianPoleI}
\end{eqnarray}
and the third at,
\begin{eqnarray}
z &=& 0\,,\nn\\
z &=& -\xib'_4\,.
\label{ClassCJacobianPoleII}
\end{eqnarray}

Here, looking at \eqn{ClassCSolutions}, we see that
powers of the loop momentum in the numerator will induce poles at
$z=\infty$ for all solutions $\Sol_i$; at,
\begin{eqnarray}
z &=& -\frac1{\tau}\,,
\label{ClassCSupplementaryPolesI}
\end{eqnarray}
for solution $\Sol_5$; and at,
\begin{eqnarray}
z &=& -\frac{\xib'_1}{\taub}\,,
\label{ClassCSupplementaryPolesII}
\end{eqnarray}
for solution $\Sol_6$.

Once again, once the dust settles on comparing ostensibly different
solutions, one finds eight independent global poles,
with $\xi'_1=\xib'_1$, $\xi'_4=\xib'_4$,
$\zeta_1=0=\zeta_2$, and
the following values of $(\xi_2,\xi'_2,\xi_3,\xi'_3)$,
\begin{eqnarray}
\Global_1:
&&\left(0,-\xib'_1,-1,0\right)
\,,\nn\\
\Global_2:
&&\left(-1,0,0,-\xib'_4\right)
\,,\nn\\
\Global_3:
&&\left(0,-\xib'_1,0,0\right)
\,,\nn\\
\Global_4:
&&(-1,0,0,0)
\,,\label{ClassCGlobalPoles}\\
\Global_5:
&&(0,0,-1,0)
\,,\nn\\
\Global_6:
&&\left(0,0,0,-\xib'_4\right)
\,,\nn\\
\Global_7:
&&\left(0,-\frac{\xib'_1}{\taub},0,\infty\right)
\,,\nn\\
\Global_8:
&&\left(-\frac{1}{\tau },0,\infty,0\right)
\,.\nn
\end{eqnarray}

Once again, for the global poles with an infinite coefficient, we must take
the limit carefully;
for global pole $\Global_7$, we should set,
\begin{equation}
\xi _2= 0
\,,\, \xi _3= 0
\,,\, \xi'_3= -\frac{\left(\xib'_1+z\right) \xib'_4}{\taub z+\xib'_1}
\,,\, \xi'_2= z
\,,\,
\end{equation}
and then take the limit $z\rightarrow -{\xib'_1}/{\taub}$,
while
for global pole $\Global_8$, we should set,
\begin{equation}
\xi'_2= 0
\,,\, \xi'_3= 0
\,,\, \xi _3= -\frac{z+1}{\tau  z+1}
\,,\, \xi _2= z
\,,\,
\end{equation}
and then take the limit $z\rightarrow -{1}/{\tau }$.

The solutions and global poles are depicted graphically in
\fig{ClassCSolutionsFigure}.

\section{Constraint Equations}
\label{ConstraintSection}

\def\jig#1{J_{\oint_{}^{}\!,#1}}
As we have seen in the previous section,
each double-box integral with a general numerator
has eight independent global poles.  One might therefore imagine obtaining
eight independent equations for coefficients of integrals by integrating
both the left- and right-hand sides of the two-loop version
of \eqn{BasicEquation} on a $T^8$ contour surrounding each global pole
in turn,
\begin{equation}
\oint_{T^8(\Global_i)} d^8v_{a,b}\; \jig{i}\,{\rm Amplitude\ Integrand}
= \sum_{j\in {\rm Basis}} c_j
\oint_{T^8(\Global_i)} d^8v_{a,b}\; \jig{i}\,{\rm Integrand}_j\,.
\end{equation}
where $T^8(\Global_i)$ denotes an eight-torus encircling the $i$th
global pole, and where $\jig{i}$ is the corresponding
Jacobian derived in the previous section.  The $v_{a,b}$, as before,
denote the eight variables $\zeta_{1,2}$, $\xi_{2,3}$, and $\xi'_{1,2,3,4}$.

  These equations would not all be valid, however, because the
derivation of the reduced form on the right-hand side assumes that
a number of integrals, with certain kinds of numerators, vanish identically.
We must restrict ourselves to contours, whether encircling one global
pole or multiple ones, that enforce the vanishing of these integrals.  As
illustrated in detail
in a simple one-loop example in ref.~\cite{MaximalTwoLoopUnitarity},
non-vanishing numerators which yield a vanishing integral over the
standard real contour will generically not vanish on a
linear combination of contours encircling
global poles.

Before deriving constraints equations for these contours,
we must address a subtlety regarding the orientation
of the eight tori $T^8(\Global_i)$.  The global poles $\Global_1, \ldots, \Global_4$ in class (b) as well as
the global poles $\Global_1, \ldots, \Global_6$ in class (c) are shared
between two distinct Riemann spheres (see figs.~\ref{ClassBSolutionsFigure}~and~\ref{ClassCSolutionsFigure}).
We adopt the convention that the corresponding residue is to be
evaluated on the sphere located towards the anti-clockwise direction of the figures. Thus, in class (b), for example, the residue at $\Global_1$
should be evaluated from $\Sol_1$; the residue at $\Global_2$ from $\Sol_3$, etc.
Moreover, we choose the orientations on each Riemann sphere such that
for any global pole $\Global_k \in \Sol_i \cap \Sol_j$, the residues
evaluated from spheres $\Sol_i$ and $\Sol_j$ are equal and opposite. That is,
for an arbitrary function~$f$ of the loop momenta one has,
\begin{equation}
\mathrm{Res}_{\Global_k} \ji f(\ell_1, \ell_2) \Big|_{\Sol_i}
= -\mathrm{Res}_{\Global_k} \ji f(\ell_1, \ell_2) \Big|_{\Sol_j} \, ,
\end{equation}
in agreement with the conventions of ref.~\cite{Caron-HuotLarsen}.
Of course, other choices of conventions are possible, but all will lead to
the same final results for the two-loop integral coefficients.

Let us now turn to the derivation of constraint equations.
We focus on the double-box topology.  Denote the insertion of
the polynomial $f(\ell_1,\ell_2)$ into the numerator of
the double box by,
\begin{equation}
\DBox[f(\ell_1,\ell_2)] =
\int {d^{D} \ell_1\over (2\pi)^D} \hspace{0.5mm} {d^{D} \ell_2\over(2\pi)^D}
\hspace{0.7mm} \frac{f(\ell_1,\ell_2)}{\ell_1^2(\ell_1 - k_1)^2
(\ell_1 - K_{12})^2(\ell_1 + \ell_2)^2\ell_2^2
(\ell_2 - k_4)^2(\ell_2 - K_{34})^2} \, .
\label{DoubleBoxWithInsertion}
\end{equation}

For class~(c), there are
two master integrals, which we take to be
\begin{equation}
I_1 = \DBox[1]\,,\quad{\rm\ and\/}\quad~
I_2 = \DBox[\ell_{1}\cdot k_{4}]\,.
\label{ClassCMasterIntegrals}
\end{equation}
For class~(b), there are three master integrals, which we take to be,
\begin{equation}
I_1 = \DBox[1]\,,\quad
I_2 = \DBox[\ell_{1}\cdot k_{4}]\,,\quad {\rm\ and\/}\quad~
I_3 = \DBox[\ell_{2}\cdot k_{1}]\,.
\label{ClassBMasterIntegrals}
\end{equation}

Vanishing integrals with non-vanishing numerators can all be thought of
as integrals of total derivatives.  They fall into two distinct classes,
parity-odd and parity-even.  The former consist of
five integrals with insertions of Levi-Civita symbols,
\begin{eqnarray}
&& \DBox[\varepsilon(\ell_1,k_2,k_3,k_4)]\,,\quad
\DBox[\varepsilon(\ell_2,k_2,k_3,k_4)]\,,\quad
\DBox[\varepsilon(\ell_1,\ell_2,k_1,k_2)]\,,\quad\nn\\
&&
\DBox[\varepsilon(\ell_1,\ell_2,k_1,k_3)]\,,\quad
\mathrm{and}\quad\DBox[\varepsilon(\ell_1,\ell_2,k_2,k_3)]\,,\quad
\label{LeviCivitaConstraints}
\end{eqnarray}
whose vanishing on our chosen contours imposes one set of constraints.

The parity-even total derivatives are precisely the integrands
which yield the equations
required to fully reduce all possible integrals arising in
a gauge theory to the two or three master integrals.

Imagine computing a two-loop amplitude with four external momenta
diagrammatically.
At the first step, the numerator of the two-loop double box will contain
dot products of the two loop momenta $\ell_{1,2}$ with external polarization
vectors, external momenta, and external spinor strings (if any external
legs are fermions).  We can express all dot products
of loop momenta with external vectors in terms of eight dot products:
$\ell_j\cdot k_1$, $\ell_j\cdot k_2$, $\ell_j\cdot k_4$, and
$\ell_j\cdot v$, where $v^\mu=\varepsilon(\mu,k_1,k_2,k_4)$.
Just as at one loop,
odd powers of $v$ will give rise to vanishing integrals,
as expressed in the Levi-Civita constraints discussed above. Even
powers can again be re-expressed in terms of the other dot products
(up to terms involving the $(-2\eps)$-dimensional components of
the loop momentum).  All integrals can then be rewritten in terms
of the six dot products of the loop momenta with
the external momenta.

Of these six dot products, three of them --- $\ell_1\cdot k_1$,
$\ell_1\cdot k_2$, $\ell_2\cdot k_4$ ---
can be rewritten as linear combinations of the propagator denominators
and external invariants.  One additional dot product of $\ell_2$ --- say
$\ell_2\cdot k_2$ --- can be rewritten in terms of the remaining two
($\ell_1\cdot k_4$ and $\ell_2\cdot k_1$), propagator denominators, and
external invariants.  The remaining two dot products are called
irreducible.

At first glance, all integrals of the form,
\begin{equation}
\DBox[( \ell_{2}\cdot k_{1})^m\,\,( \ell_{1}\cdot k_{4})^n]\,,
\end{equation}
would appear to be irreducible.  In a general gauge theory, all integrals
with $0\leq~m~+~n \leq~6$ and $m,n\leq 4$ can appear, giving a total of 22
integrals.  At a first stage, then, before using integration-by-parts (IBP) identities,
we can reduce an arbitrary double-box integral appearing in a gauge-theory
amplitude to a linear combination of the 22 different integrals that
can arise with powers of the two irreducible numerators.
 The IBP relations will allow us to reduce the integrals to
a small set of master integrals.
The number of IBP relations depends on the configuration of external
momenta.  For class~(b), we have 19 such relations, leaving the
three masters in \eqn{ClassBMasterIntegrals}.
For class~(c), we have 20 such relations, leaving us with the
two masters in \eqn{ClassCMasterIntegrals}.  Each IBP relation gives
us a nontrivial numerator which integrates to zero; explicit
examples may be found in ref.~\cite{MaximalTwoLoopUnitarity}.

In order to ensure that these equations continue to hold after
performing the global contour integrals, we seek a linear combination
of $T^8$ contours surrounding the different global poles that satisfies
\begin{equation}
\sum_{j=1}^8 a_j \oint_{T^8(\Global_j)} d^8v\;\jig{j} F = 0\,,
\end{equation}
and replace $F$ in turn by each of the
parity-odd or -even numerators discussed above.  This gives us
five equations for the parity-odd numerators, and either 19 or 20
for the parity-even ones.

It turns out that one of the five Levi-Civita equations is linearly dependent
on the others, just as in the purely massless case considered in
ref.~\cite{MaximalTwoLoopUnitarity}.  This leaves us with four
independent constraints; in classes~(b) and (c), these have the form\footnote{We remark that,
in class (b), swapping $\gamma_{12}^+ \longleftrightarrow \gamma_{12}^-$
amounts to changing the coordinate value at the location of the global poles
as well as changing the value of the corresponding residue. The constraint
equations~(\ref{eq:LeviCivitaConstraintsClassesBandC}) and (\ref{eq:IBPConstraintsClassB})
are invariant.},
\begin{equation}
% checked same for 0m, 1m, 2md, 2ml and 2ms & 3m
\left(
\begin{array}{cccccccc}
  1 & -1 &  0 &  0 &  0 &  0 &  0 &  0 \\
  0 &  0 &  1 & -1 &  0 &  0 &  0 &  0 \\
  0 &  0 &  0 &  0 &  1 & -1 &  0 &  0 \\
  0 &  0 &  0 &  0 &  0 &  0 &  1 & -1 \\
\end{array}
                  \right)\,
\left(\matrix{a_1\cr a_2\cr a_3\cr a_4\cr a_5\cr a_6\cr a_7\cr a_8\cr}\right)
= 0\,.
\label{eq:LeviCivitaConstraintsClassesBandC}
\end{equation}

Similarly, only one IBP equation is independent for class~(b),
\begin{equation}
% checked same for 2ms & 3m
\left(
\begin{array}{cccccccc}
 1 & 1 & -1 & -1 & 1 & 1 & 0 & 0 \\
\end{array}
                  \right)\,
\left(\matrix{a_1\cr a_2\cr a_3\cr a_4\cr a_5\cr a_6\cr a_7\cr a_8\cr}\right)
= 0\,;\label{eq:IBPConstraintsClassB}
\end{equation}
and
only two for class~(c),
\begin{equation}
% checked same for 0m, 1m, 2md and 2ml
\left(
\begin{array}{cccccccc}
  1 &  1 & -1 & -1 &  0 &  0 &  0 &  0 \\
  1 &  1 &  0 &  0 & -1 & -1 & -1 & -1 \\
\end{array}
                  \right)\,
\left(\matrix{a_1\cr a_2\cr a_3\cr a_4\cr a_5\cr a_6\cr a_7\cr a_8\cr}\right)
= 0\,.\label{eq:IBPConstraintsClassC}
\end{equation}

The simultaneous solutions to these equations are choices of
contours suitable for integrating on both sides of the two-loop version of \eqn{BasicEquation}
in order to derive useful relations for coefficients.

\section{Projectors}
\label{ProjectorsSection}

\def\iV#1{| #1 \rangle}
\def\iU#1{| #1 ]}
\def\Vi#1{\langle #1 |}
\def\Ui#1{[ #1 |}

\def\Integrand{{\cal I}}
\def\treelevel{{(0)}}
Imposing the constraints obtained in the previous section will leave
us with three nontrivial linearly independent combinations of contours
for external kinematics in class~(b), and two combinations in class~(c).
Using these contours on the two-loop version of \eqn{BasicEquation}
restricted to four-point amplitudes will then give us three or two
equations, respectively, for the coefficients of the three or two different
master integrals~(\ref{ClassBMasterIntegrals})
 and~(\ref{ClassCMasterIntegrals}).  We can diagonalize these
equations by requiring in turn the vanishing of each master integral
along the contour.  The resulting equations for the coefficients take
the form,
\begin{equation}
c_i = \sum_{j=1}^8 a_j \oint_{T^8(\Global_j)} d^8v_{a,b}\;\jig{j}
\sum_{{\rm particles\atop\rm helicities}}
 \prod_{p=1}^6 A^\treelevel_p(v_{a,b})\,,
\label{BasicCoefficientEquation}
\end{equation}
where the product in the integrand is over the tree amplitudes at
all vertices, with their dependence coming through the
parametrization~(\ref{SpinorParametrization})-(\ref{BasicSpinorDefinitions}).  The product is then
summed over all allowed choices of internal particles and helicities.

This abstract form of a multidimensional contour
integration hides a subtlety whose full exploration we postpone
to a future paper.  In multivariate complex contour integration, the result
is not necessarily independent of the order of integration.  In our
context, this is connected with accidental sharing of singularities between
different integrals, a feature not present in the one-loop case.  Once
we specify the ordering, the accidental sharing disappears, as does any
ambiguity in the result.  In the present context, the order
in \eqn{BasicCoefficientEquation} is specified
by performing the $z$ integration last,
\begin{equation}
\oint_{T^8(\Global_j)} d^8v\; \ji \Integrand \equiv
\oint_{C_\delta(z^0_j)} dz\; \jig{j} \Integrand\biggl|_{\Sol(\Global_j)}
\label{GlobalResidueEvaluation}
\end{equation}
where the integrand is first to be evaluated at a heptacut solution
$\Sol(\Global_j)$ containing the global pole, and the contour integral
in $z$ is then to be performed on a small circle $C_\delta$
surrounding the value of $z$ at the given global pole. In this
equation as well as in eq.~(\ref{BasicCoefficientEquation}), $\jig{j}$
refers to the Jacobian of the chosen heptacut solution.  (As remarked
at the beginning of Section~\ref{ConstraintSection}, there is more
than one heptacut solution containing a given global pole; we refer to
the discussion there for our conventions.)

In class~(b), the residues for the three master integrals are given by
\bea
I_1:&& R_1=(1,1,1,1,0,0,0,0)\,, \nn \\
I_2:&& R_2=\frac{\gamma_{12}(s_{12} s_{14}-m_1^2 m_3^2)}{2(\gamma_{12}^2-m_1^2 m_2^2)} (0,0,0,0,1,1,0,0)\,, \label{eq:Residues_of_masters_Class_b} \\
I_3:&& R_3= \frac{s_{12} (s_{14} - m_1^2)}{2 \gamma_{34}} (1, 1, 1, 1, 0, 0, 0, 0) + \frac{m_1^2 m_3^2-s_{12} s_{14}}{2 \gamma_{34}} (1, 1, 0, 0, 0, 0, 1, 1) \nn\,.
\eea

Imposing the constraints~(\ref{eq:LeviCivitaConstraintsClassesBandC})
and~(\ref{eq:IBPConstraintsClassB})
and then inverting the matrix of these residue vectors, we obtain projectors
(or master contours) isolating a single master integral.  Each is specified
by a vector giving the values of the $a_i$ to be used in the integrations
above.
In class~(b), the projectors for the three masters are,
\begin{eqnarray}
I_1:&&
P_1 = \frac14\left(1,1,1,1,0,0,1-\frac{2 m_1^2 \left(m_3^2-s_{12}\right)}{m_1^2 m_3^2-s_{12} s_{14}},1-\frac{2 m_1^2 \left(m_3^2-s_{12}\right)}{m_1^2 m_3^2-s_{12} s_{14}}\right)
\,,
\nn\\
I_2:&&
% checked same for 2ms & 3m
P_2 = \frac{N_2^{({\rm b})}}2\left(-1,-1,1,1,2,2,1,1\right)
\,,
\label{MasterFormulaeClassB}\\
I_3:&&
% checked same for 2ms & 3m
P_3 = \frac{N_3^{({\rm b})}}2\left(0,0,0,0,0,0,-1,-1\right)
\,,
\nn
\end{eqnarray}
where\footnote{For the normalization factor $N_2^{({\rm b})}$ in the
  projector for integral $I_2$, swapping $\gamma_{12}^+
  \longleftrightarrow \gamma_{12}^-$ will induce a change of sign in
  the value of $N_2^{({\rm b})}$ and the heptacut of $I_2$.},
\begin{eqnarray}
N_2^{({\rm b})} &=& \frac{\gamma_{12}^2-m_1^2 m_2^2}
              {\gamma_{12}(s_{12} s_{14}-m_1^2 m_3^2)}\,,\nn\\
N_3^{({\rm b})} &=& \frac{2 (s_{12}-m_3^2)}
              {s_{12} s_{14}-m_1^2 m_3^2}\,.
\label{ClassBProjectorNormalizations}
\end{eqnarray}
The formul\ae{} in previous sections allow for any pattern of masses
within class~(b); but the residues in eq.~(\ref{eq:Residues_of_masters_Class_b})
and the projectors here are given for the special case
that $m_4=0$, $m_1\neq 0\neq m_2$, and hold whether $m_3$ is zero or not.

In class~(c), the residues for the two master integrals are given by
\bea
I_1:&& R_1=(1,1,1,1,1,1,0,0)\,, \label{eq:Residues_of_masters_Class_c} \\
%I_2:&& R_2=-\frac{1}{2}m_4^2(1,1,1,1,0,0,0,0)+\frac{(\gamma_{12} k_1\cdot m_4- k_1^2 k_2 \cdot k_4) (\gamma_{12} + k_2^2)}{\gamma_{12}^2}(0,0,0,0,1,1,0,0)\,.\\
I_2:&& R_2=-\frac{1}{2} \left(m_4^2,m_4^2,m_4^2,m_4^2,-2k_1^\flat \cdot k_4 \Big(1 + \frac{m_2^2}{\gamma_{12}}\Big),-2k_1^\flat \cdot k_4 \Big(1 + \frac{m_2^2}{\gamma_{12}}\Big),0,0 \right) \nn \,.
\eea

In class~(c), the projectors are
\begin{eqnarray}
I_1:&&
% checked same for 0m, 1m, & 2md in m_4 -> 0 limit
P_1 = \frac{N_1^{({\rm c})}}4\left(1,1,1,1,
\frac{2 m_4^2 \left(s_{12}-m_1^2\right)}
     {s_{12} \left(s_{14}-m_4^2\right)+m_1^2 m_4^2},
\frac{2 m_4^2 \left(s_{12}-m_1^2\right)}
     {s_{12} \left(s_{14}-m_4^2\right)+m_1^2 m_4^2},
\right.\nn\\
&&\hskip 5mm\left.
1-\frac{2 m_4^2 \left(s_{12}-m_1^2\right)}
     {s_{12} \left(s_{14}-m_4^2\right)+m_1^2 m_4^2},
1-\frac{2 m_4^2 \left(s_{12}-m_1^2\right)}
     {s_{12} \left(s_{14}-m_4^2\right)+m_1^2 m_4^2}
\right)
\,,
\label{MasterFormulaeClassC}\\
I_2:&&
% checked same for 0m, 1m, 2md and 2ml
P_2 = \frac{N_2^{({\rm c})}}4\left(1,1,1,1,-2,-2,3,3\right)
\,,\nn
\end{eqnarray}
where
\begin{eqnarray}
N_1^{({\rm c})} &=& 1 + \frac{m_4^2 (m_1^2-s_{12})}{s_{12} s_{14}}\,,\nn\\
N_2^{({\rm c})} &=& -\frac1{k_{1}^\proj\cdot k_4}
\left(1 + \frac{m_4^2 (m_1^2-s_{12})}{s_{12} s_{14}}\right)\,.
\end{eqnarray}
As above,
the formul\ae{} in previous sections and in
eq.~(\ref{eq:Residues_of_masters_Class_c})
allow for any pattern of masses
within class~(c); but the projectors here are given for the special case
that $m_2=0$, $m_1\neq 0$, and at most one of $m_3$ and $m_4$ is non-zero.

\section{Examples: Higgs Amplitudes and Form Factors}
\label{ExampleSection}

In this section we will work out cut expressions, residues and integral
coefficients for the one-mass double-box integral in a few simple but
nontrivial four-point examples. We use these examples to
illustrate how to apply the method described in previous sections.
In particular we discuss a single massive scalar amplitude and a form factor. A simple way to introduce
a massive scalar in a gauge invariant way is to add the following
interaction to the pure Yang-Mills theory:
\be
{\cal L}_{\rm int.} = -\frac1{4\Lambda} H {\rm Tr}(F^{\mu
  \nu}F_{\mu \nu}) \,,
\ee
where $\Lambda$ denotes the scale at which the physics underlying this
higher-dimension operator has been integrated out, and $H$ is a
massive scalar
boson.
This is the simplest gluon--scalar interaction
for a colorless scalar not carrying a conserved charge.
It is precisely the form of the effective
Lagrangian~\cite{EffectiveHiggsLagrangian}
 for the Standard Model Higgs--gluon
interaction, once the top quark has been integrated out.
 Recently, Gehrmann {\it et al.\/} have
computed the two-loop $H\rightarrow ggg$
amplitude in this effective theory~\cite{Gehrmann},
later simplified by Duhr~\cite{Duhr}.

\def\Operator{{\cal O}}
The computation of the scalar's three-gluon decay amplitude can also be viewed
as the computation of the three-gluon form factor for the operator,
\be
\Operator_g \equiv -\frac1{4}{\rm Tr}(F^{\mu \nu}F_{\mu \nu})\,.
\ee
We can compute such form factors using the same approach as for on-shell
amplitudes,
\be
\langle0|\Operator(q)|1,2,\ldots,n\rangle =
A^{\Operator}(q,1,2,\ldots,n)\,.
\ee
In addition to the operator $\Operator_g$, we will also consider
a well-studied scalar operator in ${\cal N}=4$ super-Yang--Mills theory (SYM),
\be
\Operator_s \equiv \frac{1}{2} {\rm Tr}(\phi_{12}\phi_{12}),	
\ee
where the subscripts refer to $R$-charge indices. These indices will not
play a role here; one can think of $\phi_{12}$ as simply a complex
scalar with a conserved $U(1)$ charge:
$\phi^{12} \equiv \phi^+ = \phi$ and
$\phi_{12} \equiv \phi^- = \bar \phi$.
Brandhuber, Travaglini and Yang recently computed the three-particle
form factor
of this operator at two loops~\cite{TwoLoopFormFactors}.
We compare
our calculation of the double-box coefficients to their results.

For our computations, the only tree-level ingredients we need
are the three-point amplitudes and two-particle form
factors. The nonvanishing form factors, after stripping off color factors,
are,
\be
\langle1^+_g ,2^+_g |\Operator_g|0\rangle =
\spbsh{1}.{2}^2\,,~~~~	
\langle1^-_g ,2^-_g |\Operator_g|0\rangle = \spash{1}.{2}^2\,,~~~~
\langle1_s^{+},2_s^{+}|\Operator_s|0\rangle = 1\,.
\ee
(We omit an overall dimensionful factor of $1/\Lambda$.)

%
%%%%%%%%% FIGURE %%%%%%%%%%%%%%%%%%
\begin{figure}[t]
\begin{center}
\includegraphics[width=0.75\textwidth]{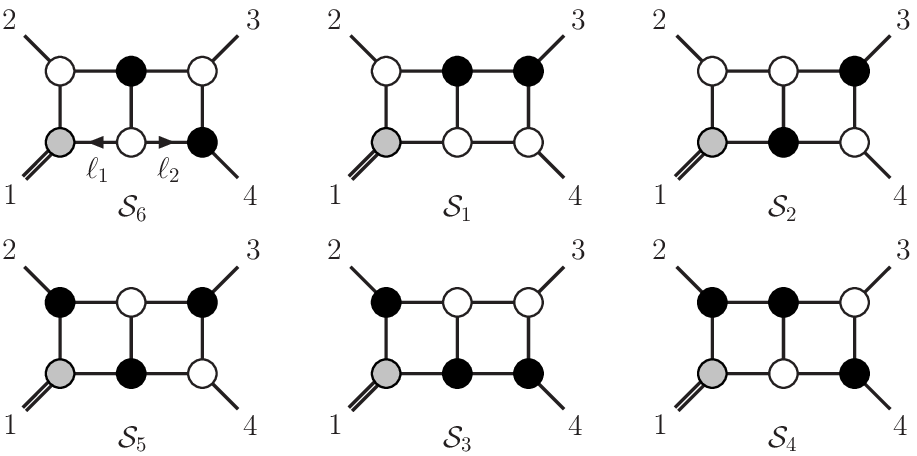}
\end{center}
\caption{\small The six solutions to the heptacut kinematics of a one-mass
  double box. White and black blobs indicate chiral and antichiral
  kinematics at the vertices, respectively.
  The gray blobs denote vertices with no definite chirality.}
\label{FormFactorCutsFigure}
\end{figure}
%%%%%%%%%%%%%%%%%%%%%%%%%%%%%%%%
%

\subsection{Computing the integral coefficients}

Let us follow the procedure derived in previous sections.
First we form the product of amplitudes
appearing in \eqn{BasicCoefficientEquation}, evaluated at
each of the heptacut solutions $\Sol_1,\ldots,\Sol_6$,
depicted in \fig{FormFactorCutsFigure}.  This is the
integrand of our remaining contour integral.
This corresponds to evaluating the factor
$\Integrand$ in \eqn{GlobalResidueEvaluation} at each of the solutions $\Sol_i$.

Next, we evaluate the remaining contour integral in $z$ by residues.
Here, we should distinguish between two types of poles: those arising
from the heptacut
Jacobian~(\ref{ClassCJacobianPoleIII})--(\ref{ClassCJacobianPoleII});
and those arising from powers of the loop momentum in the integrand
itself~(\ref{ClassCSupplementaryPolesI})--(\ref{ClassCSupplementaryPolesII}).
The integrand can safely be evaluated
at the former type, where it is finite.

\def\MSYM{${\cal N}=4$ SYM}
The poles arising from powers of the loop momentum are at infinite
values of one of the loop momenta. (The poles are
nonetheless at finite values of the
remaining integration variable $z$.)
The Jacobian factor remains finite at these poles.
Here the procedure is more
involved, because in general the poles may be of higher order.
The order of the pole depends on the ultraviolet power-counting behavior
of the process in the given theory.
In \MSYM, and for the form factors we consider here, the
poles will be simple.
A general fermion and scalar content in the theory would lead
to poles of higher order.
(From a computational point of view, the residues at
infinite loop momentum require a more careful treatment as they may receive
contributions that arise from the interference between singular
and finite terms in the Laurent expansion of the various loop-momentum
factors entering the integrand.)

\def\Residue{R}
For each kinematical solution $\Sol_j$ we denote the integrand by
$\Integrand_j$,
\be \Integrand_j \, \equiv \,
(-i)^7 \sum_{{\rm particles\atop\rm helicities}}
\prod_{i=1}^6 A^\treelevel_{i}(\ell_1,\ell_2)\Big|_{{\cal
    S}_j}\,.
\ee
At each global pole, the remaining integrand in $z$ has a pole at
a single location $z^0_j$; we evaluate the residue at that pole,
\be
\Residue_{{\cal G}_j} = \Res_{z=z^0_j} \jig{j} \Integrand({\cal G}_j)\,,
\ee
as described above.

%
%%%%%%%%% FIGURE %%%%%%%%%%%%%%%%%%
\begin{figure}[t]
\begin{center}
\includegraphics[width=0.75\textwidth]{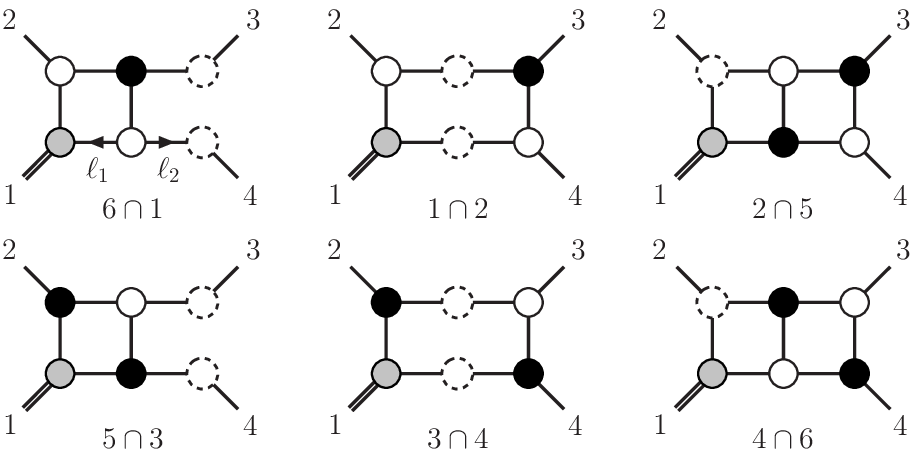}
\end{center}
\caption{\small The Jacobian poles are located at the intersection
  points of the six kinematical solutions, denoted by $i\cap j$ for solutions $\Sol_i$ and $\Sol_j$. They correspond to situations where vertical rungs become soft (in the massless case),
  as indicated by the absent lines, or in a more general situation
  where the kinematics becomes collinear, as indicated by the dashed
  blobs.}
\label{FormFactorJacobianPolesFigure}
\end{figure}
%%%%%%%%%%%%%%%%%%%%%%%%%%%%%%%%
%

\def\MHVbar{$\overline{\rm MHV}$}
The results below are valid for any massless gauge theory with $n_f$
fermions and $n_s$ (complex) scalars, both in the adjoint
representation. The kinematical solutions and poles in the one-mass
case are illustrated in \fig{FormFactorCutsFigure} and
\fig{FormFactorJacobianPolesFigure}, respectively.
We can read off the amplitudes from the pattern of white and black
blobs in the figures.
Maximally helicity violating (MHV)
 amplitudes are non-vanishing for chiral kinematics, but vanish
for antichiral kinematics.  The white blobs therefore give rise to
MHV amplitudes.  Similarly,
\MHVbar{} amplitudes are non-vanishing for antichiral kinematics,
but vanish for chiral kinematics.  The black blobs
therefore give rise to \MHVbar{} amplitudes.
More detailed
information on notation and the location of the poles in the one-mass
case is summarized in \tab{PolesTable}.

%%%%%%%% TABLE %%%%%%%%%%%
\def\hs{\hskip .2 cm \null }
\begin{table*}[t]
\caption{The properties of the global poles in the one-mass case:
  Here we assume that $m_2=m_3=m_4=0$. The notation $i\cap j$
  indicates that the pole is located at the intersection point of the
  Riemann spheres labeled by $\Sol_i$ and $\Sol_j$, and $\infty_{iR}$
  corresponds to poles in $\Sol_i$ at infinite $\ell_2$ momentum (see
  ref.~\cite{Caron-HuotLarsen}).}
\label{PolesTable} 	
\vskip .4 cm
\begin{tabular}{|c|c|c|c|c|c|}
\hline
pole 	& location & $(\xi_2,\xi'_2,\xi_3,\xi'_3)$ & $\ell_1$ & $\ell_2$ \\
\hline
${\cal G}_{1}$  & $1\cap2$ & $(0,-1,-1,0)$ & $-\frac{1}{2}\Ui{4} \gamma^\mu \iV{1^\flat}\frac{\spash{4}.{3}}{\spash{1^\flat}.{3}}$ & $\frac{1}{2}\Ui{4} \gamma^\mu \iV{1^\flat}\frac{\spash{4}.{3}}{\spash{1^\flat}.{3}}$   \\
\hline
${\cal G}_{2}$  & $3\cap4$  & $(-1,0,0,-1)$ & $-\frac{1}{2}\Vi{4} \gamma^\mu \iU{1^\flat}\frac{\spbsh{4}.{3}}{\spbsh{1^\flat}.{3}}$ & $\frac{1}{2}\Vi{4} \gamma^\mu \iU{1^\flat}\frac{\spbsh{4}.{3}}{\spbsh{1^\flat}.{3}}$   \\
\hline
${\cal G}_{3}$  & $6\cap1$ & $(0,-1,0,0)$ & $-\frac{1}{2}\Ui{4} \gamma^\mu \iV{1^\flat}\frac{\spash{4}.{3}}{\spash{1^\flat}.{3}}$ & $k_4$ \\
\hline
${\cal G}_{4}$  & $5\cap3$ & $(-1,0,0,0)$ & $-\frac{1}{2}\Vi{4} \gamma^\mu \iU{1^\flat}\frac{\spbsh{4}.{3}}{\spbsh{1^\flat}.{3}}$ & $k_4$   \\
\hline
${\cal G}_{5}$  & $2\cap5$ & $(0,0,-1,0)$ & $k_1^\flat$ & $\frac{1}{2}\Ui{4} \gamma^\mu \iV{1^\flat}\frac{\spash{4}.{3}}{\spash{1^\flat}.{3}}$  \\
\hline
${\cal G}_{6}$  & $4\cap6$ & $(0,0,0,-1)$ & $k_1^\flat$ & $\frac{1}{2}\Vi{4} \gamma^\mu \iU{1^\flat}\frac{\spbsh{4}.{3}}{\spbsh{1^\flat}.{3}}$  \\
\hline
${\cal G}_{7}$  & $\infty_{6R}$ & $\text{\Large(}0,-\frac{\spbsh{3}.{1^\flat}\spbsh{4}.{2}}{\spbsh{3}.{2}\spbsh{4}.{1^\flat}},0,\infty\text{\Large)}$ & $-\frac{1}{2}\Ui{3} \gamma^\mu \iV{1^\flat}\frac{\spash{3}.{4}}{\spash{1^\flat}.{4}}$ & $\infty \, \Ui{3} \gamma^\mu \iV{4}$  \\
\hline
${\cal G}_{8}$  & $\infty_{5R}$ & $\text{\Large(}\!-\frac{\spash{3}.{1^\flat}\spash{4}.{2}}{\spash{3}.{2}\spash{4}.{1^\flat}},0,\infty,0\text{\Large)}$ & $-\frac{1}{2}\Vi{3} \gamma^\mu \iU{1^\flat}\frac{\spbsh{3}.{4}}{\spbsh{1^\flat}.{4}}$ & $\infty \,  \Vi{3} \gamma^\mu \iU{4}$ \\
\hline
\end{tabular}
\vskip .5 cm
\end{table*}
%%%%%%%%%%%%%%%%%%%%%%%

\subsubsection{The $(\Operator_g,2^-,3^-,4^-)$ Configuration}

It will be helpful to label the internal cut lines as follows,
\be
\ell_3=\ell_1-k_1\,,~\ell_4=\ell_1-k_1-k_2\,,~\ell_5=\ell_2-k_4\,,
~\ell_6=\ell_2-k_4-k_3\,,~\ell_7=-\ell_1-\ell_2\,.
\ee
For each heptacut solution,
compute the cut integrand by multiplying together
tree amplitudes for each vertex,
and summing over different possible assignments of
internal states.
For example, for the evaluation on $\Sol_6$,
here only a single configuration of internal helicity assignments contributes,
\bea
\Integrand_{6}
&=&(-i)^7A^{\rm Higgs}(1^{\rm H},\ell_3^{-},-\ell_1^{-})
A^\treelevel(2^{-},\ell_4^{-},-\ell_3^{+})
A^\treelevel(3^{-},-\ell_5^{-},\ell_6^{+})
A^\treelevel(4^{-},-\ell_2^{+},\ell_5^{+}) \nn \\
&&\times
A^\treelevel(\ell_1^{+},\ell_7^{-},\ell_2^{-})
A^\treelevel(-\ell_4^{+},-\ell_6^{-},-\ell_7^{+}) \nn \\
&=&-\spash{\ell_1}.{ \ell_3}^2
\frac{\spash{2}.{\ell_4}^3}{\spash{\ell_4}.{\ell_3} \spash{\ell_3}.{2}}
\frac{\spash{3}.{\ell_5}^3}{\spash{\ell_5}.{\ell_6}\spash{\ell_6}.{3}}
\frac{\spbsh{\ell_2}.{\ell_5}^3}{\spbsh{\ell_5}.{4}\spbsh{4}.{\ell_2}}
\frac{\spash{\ell_7}.{\ell_2}^3}{\spash{\ell_2}.{\ell_1} \spash{\ell_1}.{\ell_7}}
\frac{\spbsh{\ell_7}.{\ell_4}^3}{\spbsh{\ell_4}.{\ell_6} \spbsh{\ell_6}.{\ell_7}}\nn \\
&=&m_1^4\frac{\ssandp2.{1\ell_1}.4
%\langle 2 | 1 \ell_1 | 4\rangle
 \spash{4}.{3}}{\spbsh{2}.{3}}\,,
\eea
where the last line is obtained after using some spinor manipulations.

The integrand takes the same simple form on $\Sol_1$
and $\Sol_2$, whereas it vanishes identically on $\Sol_5$,
$\Sol_3$ and $\Sol_4$, that is
\be
\Integrand_{6}=\Integrand_{1}=\Integrand_{2}=m_1^4
\frac{\ssandp2.{1\ell_1}.4
%\langle 2 | 1 \ell_1 | 4\rangle
\spash{4}.{3}}{\spbsh{2}.{3}}\,,~~\Integrand_{5}=\Integrand_{3}=\Integrand_{4}=0\,.
\label{firstconfiguration}
\ee
These results are the same independent of the number of scalars
or fermions in the gauge theory. In particular this means that, for
the $(\Operator_g,2^-,3^-,4^-)$ configuration, all (massless) gauge
theories have the same double-box cuts as the ${\cal N}=4$ theory,
explaining the simple dependence on loop momenta in
\eqn{firstconfiguration}.

To obtain the eight residues we should evaluate the integrand in
the neighborhood of the global poles, obtaining
\bea
\Integrand_{6,1}({\cal G}_3)&=&\Integrand_{1,2}({\cal G}_1)=m_1^4 s_{34}\frac{ \spash{2}.{3}  \spash{3}.{4}}{\spbsh{2}.{4}}\,, \nn \\
\Integrand_{6}({\cal G}_7)&=&m_1^4 s_{34} \frac{ \spash{2}.{4} \spash{4}.{3}}{\spbsh{2}.{3}}\,, \nn \\
\Integrand_{5}({\cal G}_8)&=&\Integrand_{2,5}({\cal G}_5)=\Integrand_{5,3}({\cal G}_4)=\Integrand_{3,4}({\cal G}_2)=\Integrand_{4,6}({\cal G}_6)=0\,,
\eea
where the notation $\Integrand_{i,j}$ indicates that either solution
$\Sol_i$ or solution $\Sol_j$ can be used.
The integrand happens to be finite at all global poles, so only the
Jacobian poles can give rise to nonvanishing residues. As mentioned,
the Jacobian has only simple poles at the intersection
points of the kinematical solutions.
With our normalization, there is no additional
dimensionful quantity that appears (it was removed in the transition
from $\ji^{(0)}$ to $\ji$).

Finally, after multiplying the integrand points $\Integrand_i$ by the
Jacobian factors, there are exactly two nonzero residues:
\be
R_{{\cal G}_3}=R_{{\cal G}_1}= m_1^4 s_{34}\frac{ \spash{2}.{3}  \spash{3}.{4}}{\spbsh{2}.{4}}\,.
\ee
Note that $R_{{\cal G}_7}=0$ even though the integrand
is nonvanishing at this point.
In vector notation the residues are
\be
R= m_1^4 s_{34}\frac{ \spa{2}.{3}  \spa{3}.{4}}{\spb{2}.{4}} (1,0,1,0,0,0,0,0)\,.
\ee
The integral coefficients are then given by a sum over appropriately
weighted residues. For the double box $\DBox[1]$ we have
\be
c_1=P_1\cdot R=\frac{1}{4}(R_{{\cal G}_1}+R_{{\cal G}_2}
+R_{{\cal G}_3}+R_{{\cal G}_4}
+R_{{\cal G}_7}+R_{{\cal G}_8})
= \frac{1}{2} m_1^4 s_{34}\frac{ \spash{2}.{3} \spash{3}.{4}}{\spbsh{2}.{4}}\,,
\ee
and for the double box $\DBox[\ell_{1}\cdot k_{4}]$,
\bea
c_2=P_2\cdot R&=&-\frac{1}{4 k_1^\flat \cdot k_4}(
R_{{\cal G}_1}
+R_{{\cal G}_2}
+R_{{\cal G}_3}
+R_{{\cal G}_4}
-2R_{{\cal G}_5}
-2R_{{\cal G}_6}
+3R_{{\cal G}_7}
+3R_{{\cal G}_8}
) \nn \\
&=&-\frac{1}{2} m_1^4 s_{34}\frac{ \spash{2}.{3}  \spash{3}.{4}}{\spbsh{2}.{4}}\frac{1}{k_1^\flat \cdot k_4}\,,
\eea
where $P_1$ and $P_2$ are given in eq.~(\ref{MasterFormulaeClassC}).

\subsubsection{The $(\Operator_g,2^+,3^-,4^-)$ Configuration}

For this configuration, we simply quote the main results.
The integrand evaluated on the solutions yields,
\be
\Integrand_{6}=\Integrand_{1}=\Integrand_{2}=0\,,~~ \Integrand_{5}=\Integrand_{3}=\Integrand_{4}=s_{34}\frac{\spash{3}.{4}^3 [4 | \ell_1 1 | 2]}{\spash{3}.{2}}\,,
\ee
valid for any number of fermions and scalars.

Because the integrands are at most linear in momenta the poles will be
simple. It is then straightforward to evaluate the residues in each
case by plugging in the leading momentum behavior at the pole. The
only nonvanishing residues are
\be
R_{{\cal G}_4}=R_{{\cal G}_2}= s_{34}^2
\frac{\spash{3}.{4}^3 \spbsh{3}.{2}}{\spash{2}.{4}}\,,
\ee
or alternatively
\be
R= s_{34}^2 \frac{\spa{3}.{4}^3 \spb{3}.{2}}{\spa{2}.{4}} (0,1,0,1,0,0,0,0)\,.
\ee
The integral coefficients are then given by
\be
c_1=P_1\cdot R= \frac{1}{2} s_{34}^2 \frac{\spash{3}.{4}^3 \spbsh{3}.{2}}{\spash{2}.{4}}\,,
\ee
and
\be
c_2=P_2\cdot R=-\frac{1}{2} s_{34}^2 \frac{\spash{3}.{4}^3 \spbsh{3}.{2}}{\spash{2}.{4}}\frac{1}{k_1^\flat \cdot k_4}\,.
\ee

\subsubsection{The $(\Operator_g,2^-,3^+,4^-)$ Configuration}

For this configuration, the results for the
various solutions are,
\bea
\Integrand_{6}&=&\Integrand_{1}=\Integrand_{2}=0\,,~~ \Integrand_{3}=\Integrand_{4}= [2 | 1 \ell_1 | 3]
\spbsh{3}.{4} \spash{4}.{2}^3\,, \nn \\
\Integrand_{5}&=&[2 | 1 \ell_1 | 3] \spbsh{3}.{4} \spash{4}.{2}^3+(n_f-4)
\frac{\ssand2.{\ell_1}.3\ssandpp3.{\ell_2}.4
%\langle 2 | \ell_1 | 3][3 | \ell_2 | 4\rangle
\spash{4}.{2}
%[2 | 1 \ell_1 \ell_2 | 2\rangle
\ssandpp2.{1\ell_1\ell_2}.2}{s_{34}} \nn \\
&&+2(n_s-n_f+1)\frac{
%[3 | \ell_2 | 4\rangle^2
\ssandpp3.{\ell_2}.4^2
\langle 2 | \ell_2 \ell_1 | 2\rangle
[2 | 1 \ell_1 \ell_2 | 2\rangle}{s_{34}^2}
\eea
where we have organized the $\Integrand_{5}$ expression such that first term is
the \MSYM{} contribution, the second term is the contribution
from a chiral ${\cal N}=1$ multiplet, and the last term is the scalar
contribution. Note that the fermions and scalars only propagate in the
rightmost loop, so any potential Yukawa couplings are not relevant here.

All integrands but $\Integrand_{5}$ are linear in momenta, giving simple
poles. The $\Integrand_{5}$ expression has up to fourth order poles.
Somewhat surprisingly, upon careful analysis, the residues at the simple poles
receive no contribution from the fermions and scalars.  All nonvanishing
residues are again equal,
\be
R_{{\cal G}_4}=R_{{\cal G}_2}= s_{34} \spbsh{2}.{3} \spash{4}.{2}^3 \spbsh{3}.{4}\,,
\ee
or alternatively
\be
R= s_{34} \spb{2}.{3} \spa{4}.{2}^3 \spb{3}.{4} (0,1,0,1,0,0,0,0)\,.
\ee
The integral coefficients are
\be
c_1=P_1\cdot R= \frac{1}{2}s_{34} \spbsh{2}.{3} \spash{4}.{2}^3 \spbsh{3}.{4}\,,
\ee
and
\be
c_2=P_2\cdot R=-\frac{1}{2} s_{34} \spbsh{2}.{3} \spash{4}.{2}^3 \spbsh{3}.{4}\frac{1}{k_1^\flat \cdot k_4} \,.
\ee

\subsubsection{The $(\Operator_g,2^-,3^-,4^+)$ Configuration}

For this configuration the integrands are,
\bea
\Integrand_{6}&=&\Integrand_{1}=\Integrand_{2}=0\,,~~ \Integrand_{5}=\Integrand_{4}= [2 | 1 \ell_1 | 4] \spbsh{4}.{3} \spash{3}.{2}^3\,, \nn \\
\Integrand_{3}&=&[2 | 1 \ell_1 | 4] \spbsh{4}.{3} \spash{3}.{2}^3+(n_f-4)\frac{\langle 2 | \ell_1 | 4][4 | \ell_2 | 3\rangle \spash{3}.{2} [2 | 1 \ell_1 \ell_2 | 2\rangle}{s_{34}} \nn \\
&&+2(n_s-n_f+1)\frac{[4 | \ell_2 | 3\rangle^2\langle 2 | \ell_2 \ell_1 | 2\rangle [2 | 1 \ell_1 \ell_2 | 2\rangle}{s_{34}^2}\,.
\eea
A helpful observation is that these integrands may be related directly
to integrand results for the $(\Operator_g,2^-,3^+,4^-)$ configuration, by
swapping $k_3\leftrightarrow k_4$, $\Integrand_3\leftrightarrow \Integrand_5$ (and
$\Integrand_1\leftrightarrow \Integrand_6$). This is a quite general feature as it
follows from using the BCJ~\cite{BCJ} amplitude relation
$(\ell_2-k_4)^2A^\treelevel(3,4, -\ell_2,\ell_6)=(\ell_2-k_3)^2A^\treelevel(4,3,
-\ell_2,\ell_6)$ for the product of trees that $k_3$ and $k_4$ attach
to.

As before the residues have no contribution from the fermions and
scalars, and the nonvanishing residues are equal,
\be
R_{{\cal G}_4}=R_{{\cal G}_2}= \frac{s_{23} s_{34} \spash{2}.{3}^3 \spbsh{3}.{4}}{\spash{2}.{4}}\,,
\ee
or alternatively,
\be
R= \frac{s_{23} s_{34} \spa{2}.{3}^3 \spb{3}.{4}}{\spa{2}.{4}} (0,1,0,1,0,0,0,0)\,.
\ee
The integral coefficients are,
\be
c_1=P_1\cdot R= \frac{1}{2} \frac{s_{23} s_{34} \spash{2}.{3}^3 \spbsh{3}.{4}}{\spash{2}.{4}}\,,
\ee
and
\be
c_2=P_2\cdot R=-\frac{1}{2} \frac{s_{23} s_{34} \spash{2}.{3}^3 \spbsh{3}.{4}}{\spash{2}.{4}}\frac{1}{k_1^\flat \cdot k_4}\,.
\ee

Finally, we note that the remaining nonvanishing helicity configurations $(\Operator_g, 2^+, 3^+, 4^+)$, $(\Operator_g, 2^-, 3^+, 4^+)$, $(\Operator_g, 2^+, 3^-, 4^+)$ and $(\Operator_g, 2^+, 3^+, 4^-)$
can be obtained by
conjugating spinors in the results quoted above: that is,
using $\spash{i}.{j}\leftrightarrow \spbsh{j}.{i}$.

\subsection{Scalar operator}

The scalar operator $\Operator_s = \frac{1}{2} {\rm Tr}(\phi_{12}\phi_{12})$
carries $R$-charge.  To obtain a non-zero result,
 the external states of
the form factor must carry the opposite charge (in the all-outgoing
convention). This means that the form factors involving only gluon
states vanish. At three points only the form factor with two scalars, or
one scalar and two fermions, are non-vanishing. Here we consider only
the form factors for two external scalars and a gluon. The results are
valid for any massless theory with $n_f\ge1$ fermions and $n_s\ge1$
scalars that has a Yukawa coupling $g_{\rm Yukawa} \bar
\psi^1\psi^2\phi_{12}$, and $g_{\rm Yukawa} = g_{\rm YM}$. Then all
the cuts become identical to those of ${\cal N}=4$ SYM.

\subsubsection{The $(\Operator_s,2_s^+,3_s^+,4^+)$ Configuration}
\label{Case1OsSection}

For the configuration $(\Operator_s,2_s^+,3_s^+,4^+)$ the cut integrands are
\be
\Integrand_{6}=\Integrand_{1}=\Integrand_{2}=0\,,~~~\Integrand_{5}=\Integrand_{3}=\Integrand_{4}= \langle 3 | ( \ell_1+k_4) | 2] \spash{2}.{3} \spbsh{3}.{4}^2 \,.
\ee
All the poles will be at most simple, and it is straightforward to obtain the residues in each case. The only nonvanishing ones are
\be
R_{{\cal G}_4}=R_{{\cal G}_2}= s_{23} s_{34} \frac{\spbsh{3}.{4} \spash{2}.{3}}{\spash{2}.{4}}\,,
\label{Case1res}
\ee
or, using vector notation,
\be
R= s_{23} s_{34} \frac{\spb{3}.{4} \spa{2}.{3}}{\spa{2}.{4}} (0,1,0,1,0,0,0,0)\,.
\ee
The integral coefficients are then given by
\be
c_1=P_1\cdot R=\frac{1}{2} s_{23} s_{34} \frac{\spbsh{3}.{4} \spash{2}.{3}}{\spash{2}.{4}}\,,
\ee
and
\be
c_2=P_2\cdot R=-\frac{1}{2} s_{23} s_{34} \frac{\spbsh{3}.{4} \spash{2}.{3}}{\spash{2}.{4}}\frac{1}{k_1^\flat \cdot k_4}\,.
\ee

\subsubsection{The $(\Operator_s,2_s^+,3^+,4_s^+)$ Configuration}

The case $(\Operator_s,2_s^+,3^+,4_s^+)$ is related to the above
configuration via the swap $k_3\leftrightarrow k_4$,
$\Integrand_3\leftrightarrow \Integrand_5$,
$\Integrand_1\leftrightarrow \Integrand_6$. The integrand is then
\be
\Integrand_{6}=\Integrand_{1}=\Integrand_{2}=0\,,~~~\Integrand_{5}=\Integrand_{3}=\Integrand_{4}= \langle 4 | ( \ell_1+k_3) | 2] \spash{2}.{4} \spbsh{4}.{3}^2
\,.\ee
The nonvanishing residues are again equal
\be
R_{{\cal G}_4}=R_{{\cal G}_2}=- s_{23} s_{34} \frac{\spbsh{3}.{4} \spash{2}.{4}}{\spash{2}.{3}}\,;
\ee
alternatively we have
\be
R=- s_{23} s_{34} \frac{\spb{3}.{4} \spa{2}.{4}}{\spa{2}.{3}}(0,1,0,1,0,0,0,0)\,.
\ee
The integral coefficients are then given by
\be
c_1=P_1\cdot R=-\frac{1}{2}s_{23} s_{34} \frac{\spbsh{3}.{4} \spash{2}.{4}}{\spash{2}.{3}}\,,
\ee
and
\be
c_2=P_2\cdot R=\frac{1}{2} s_{23} s_{34} \frac{\spbsh{3}.{4} \spash{2}.{4}}{\spash{2}.{3}} \frac{1}{k_1^\flat \cdot k_4}\,.
\ee

\subsubsection{The $(\Operator_s,2^+,3_s^+,4_s^+)$ Configuration}
Finally, for the configuration $(\Operator_s,2^+,3_s^+,4_s^+)$, we have
\be
\Integrand_{6}=\Integrand_{1}=\Integrand_{2}=0\,,~~~\Integrand_{5}=\Integrand_{3}=\Integrand_{4}= \frac{\langle 3 | ( \ell_1+k_4) | 2] s_{34}^2 }{\spash{3}.{2}}\,.
\ee
The nonvanishing residues are
\be
R_{{\cal G}_4}=R_{{\cal G}_2}= s_{34}^2 \frac{\spbsh{2}.{3} \spash{3}.{4}}{\spash{2}.{4}}\,;
\ee
alternatively,
\be
R= s_{34}^2 \frac{\spb{2}.{3} \spa{3}.{4}}{\spa{2}.{4}}(0,1,0,1,0,0,0,0)\,.
\ee
The integral coefficients are then given by
\be
c_1=P_1\cdot R=\frac{1}{2}s_{34}^2 \frac{\spbsh{2}.{3} \spash{3}.{4}}{\spash{2}.{4}}\,,
\ee
and
\be
c_2=P_2\cdot R=-\frac{1}{2} s_{34}^2 \frac{\spbsh{2}.{3} \spash{3}.{4}}{\spash{2}.{4}}\frac{1}{k_1^\flat \cdot k_4}\,.
\ee

The remaining nonvanishing gluon-scalar configurations $({\cal
  O}_s,2_s^+,3_s^+,4^-)$, $(\Operator_s,2_s^+,3^-,4_s^+)$ and $({\cal
  O}_s,2^-,3_s^+,4_s^+)$ can be obtained from spinor conjugations of
the above ones. (The fact that the scalars are complex does not
interfere with this relation, as has been explicitly checked.)

\subsubsection{Comparison to Previous Results for
$(\Operator_s,2_s^+,3_s^+,4^+)$}

Here we compare to the results of Brandhuber, Travaglini and Yang
\cite{TwoLoopFormFactors}. They obtain the following answer for the
numerator of the double-box integral:
\be
N_{\rm DB}=\frac{\spash{2}.{3}}{\spash{3}.{4}\spash{4}.{2}}s_{34}(s_{24}\ell_1\cdot k_4-s_{23}\ell_1\cdot k_3)\,.
\ee
The integrand of the maximal cut is simply this numerator
$\Integrand_i=N_{\rm DB}$ for all kinematical solutions. Clearly, this
integrand is different from the ones we computed above. However, the
difference should be due to terms that integrate to zero.

Computing the residues we recognize the same expression that appeared
in \eqn{Case1res}. But this time there are more nonzero residues,
\be
R_{{\cal G}_3}=R_{{\cal G}_1}=R_{{\cal G}_4}=R_{{\cal G}_2}=\frac{1}{2}s_{23} s_{34} \frac{\spbsh{3}.{4} \spash{2}.{3}}{\spash{2}.{4}}\,.
\ee
Two of the residues vanish due to an identity,
\be
R_{{\cal G}_5}=R_{{\cal G}_6}=\frac{1}{2} \frac{\spbsh{3}.{4}\spash{2}.{3}}{\spash{2}.{4}}(2s_{24}k_1^\flat \cdot k_4-2s_{23}k_1^\flat \cdot k_3)=0\,.
\ee
Writing this as a vector gives
\be
R=\frac{1}{2}s_{23} s_{34} \frac{\spb{3}.{4} \spa{2}.{3}}{\spa{2}.{4}}(1,1,1,1,0,0,0,0)\,.
\ee
The integral coefficients are then given by
\be
c_1=P_1\cdot R=\frac{1}{2} s_{23} s_{34} \frac{\spbsh{3}.{4} \spash{2}.{3}}{\spash{2}.{4}}\,,
\ee
and
\be
c_2=P_2\cdot R=-\frac{1}{2} s_{23}s_{34}\frac{\spbsh{3}.{4}\spash{2}.{3}}{\spash{2}.{4}} \frac{1}{k_1^\flat \cdot k_4}\,.
\ee

These results agree with the integral coefficients that we computed
in \sect{Case1OsSection}, so we find complete agreement with the
results of ref.~\cite{TwoLoopFormFactors}.

\section{Conclusions}
\label{ConclusionSection}

In this paper, we have continued the approach of
ref.~\cite{MaximalTwoLoopUnitarity} in developing the maximal
generalized unitarity method for two-loop amplitudes.
  Cutting
propagators can be viewed as replacing the original real loop-momentum
contours of integration by contours encircling the global poles of
the integrand.  Double-box integrals, independent of the configuration
of external masses, have eight independent global poles.  The possible
ways of cutting propagators, or equivalently the allowed linear combinations
of contours surrounding global poles, are constrained by the requirement
 that the evaluation along any contour respect the
vanishing of certain insertions of Levi-Civita symbols, as well as of
total derivatives arising from integration-by-parts identities.
We
derived the corresponding constraint equations for double
boxes with up to three external masses.
The number of
master integrals depends on the configuration of the external momenta,
but in all cases the constraint equations yield unique and simple
formul\ae{} for the coefficients of the master integrals.
The master formula~(\ref{BasicCoefficientEquation}), along with
the projectors, given in
\eqns{MasterFormulaeClassB}{MasterFormulaeClassC},
are our principal results.

\section*{Acknowledgments}

\vskip -.3 cm
We thank Simon
Caron-Huot for many helpful discussions.
This work is supported by
the European Research Council under Advanced Investigator Grant
ERC--AdG--228301.

%%%%%%%%%%%%%%%%%%%%%%%%%%%%%%%%%%%%%%%%%%

%preprint only
%\mciteSetSublistMode{n}

\end{document}